\documentclass[preprint,prd,aps,showpacs,showkeys,nofootinbib]{revtex4}
\usepackage{graphicx}
\usepackage{makecell}
\usepackage{bm}
\usepackage{url}
\begin{document}


\title{95 GeV excess in a $CP$-violating $\mu$-from-$\nu$ SSM}

\author{Chang-Xin Liu$^{a,b,c}$\footnote{LIUchangxinZ@163.com},
Yang Zhou$^{a,b,c}$,
Xiao-Yu Zheng$^{a,b,c}$, \\
Jiao Ma$^{a,b,c}$,
Tai-Fu Feng$^{a,b,c,d,e}$\footnote{fengtf@hbu.edu.cn},
Hai-Bin Zhang$^{a,b,c,d}$\footnote{hbzhang@hbu.edu.cn}}

\affiliation{$^a$Department of Physics, Hebei University, Baoding, 071002, China\\
$^b$Hebei Key Laboratory of High-precision Computation and Application of Quantum Field Theory, Baoding, 071002, China\\
$^c$Hebei Research Center of the Basic Discipline for Computational Physics, Baoding, 071002, China\\
$^d$Institute of Life Science and Green Development, Hebei University, Baoding, 071002, China\\
$^e$Department of Physics, Guangxi University, Nanning, 530004, China\\
$^f$College of Physics, Chongqing University, Chongqing, 400044, China}

\begin{abstract}
The CMS and ATLAS Collaborations have recently reported their results searching for a light Higgs boson with mass around 95 GeV, based on the full run 2 dataset. In the framework of the $CP$-violating (CPV) $\mu$-from-$\nu$ supersymmetric standard (SSM), we discuss an $\sim$ 2.9$\sigma$ (local) excess at 95 GeV in the light Higgs boson search in the diphoton decay mode as reported by ATLAS and CMS, together with an $\sim$ 2$\sigma$ excess (local) in the $b\bar{b}$ final state at LEP in the same mass range. By introducing CPV phases as well as by mixing $CP$-even Higgs and $CP$-odd Higgs, a lighter Higgs boson in the $\mu\nu$SSM can be produced, which can account for the ``diphoton excess''.

\end{abstract}

\keywords{Supersymmetry, Higgs boson decay, CP violation}
\pacs{12.60.Jv, 14.80.Da}

\maketitle

\section{Introduction\label{sec1}}
In 2012, the 125 GeV Higgs boson discovered by the Large Hadron Collider (LHC) \cite{mh-ATLAS,mh-CMS}; the measured mass of the Higgs boson now is~\cite{PDG1} $m_h=125.25\pm 0.17\: {\rm{GeV}}$. The discovery of the Higgs boson marked a huge success for the standard model (SM), but it did not stop the search for new physics (NP) at the LHC, and one of them was the search for lighter scalar particles.

Searches for a lighter Higgs have been performed at the LEP \cite{LEP0,LEP1,LEP2}, the Tevatron \cite{Tevatron} and the LHC \cite{CMS0,CMS1,CMS2,CMS3,CMS4,CMS5,CMS6,95ATLAS,CMS8}. Interestingly, the excesses observed by CMS and LEP occurred at a similar mass range \cite{CMS9}. CMS has performed searches for scalar diphoton resonances at 8 and 13 TeV \cite{CMS8,CMS9}; based on the 8 TeV data and the 13 TeV data, integrated luminosity is 19.7 and 35.9 fb$^{-1}$ respectively, which showed a 2.8 $\sigma$ local excess at 95.3 GeV \cite{CMS1,CMS5,CMS9}. Since the excess was performed \cite{CMS1,CMS5}, it has received considerable attention \cite{attention,attention1,attention2,attention3,attention4,attention5,attention6,attention7,attention8,attention9,attention10}. The excesses have been discussed in several models, in a natural next-to-minimal supersymmetric standard model (SSM) \cite{98LEP1}, and a general next-to-minimal supersymmetric standard model \cite{attention}. In Ref.\cite{new request}, the authors performed a recombination of all the signatures leading to an excess for a 95 GeV scalar candidate, and a real triplet interpretation of such excess in Ref.\cite{new request1}. Furthermore, in Ref.\cite{new request2}, the authors studied the excess in the top-antitop differential distribution carried out by ATLAS. A study of the phenomenology of the 95 GeV scalar in a UV complete model in Ref.\cite{new request3}. In Refs.\cite{98LEP3,search}, the authors have discussed the excesses in a $CP$-conserving $\mu$-from-$\nu$ SSM. In the two-Higgs doublet model with an additional real singlet, the excesses have also been discussed in Refs.\cite{search1,search2,search3}. In this work, we find a suitable parameter spaces which can explain the 95 GeV excess.

As one of the extensions of the SM, the $\mu$-from-$\nu$ supersymmetric standard model\cite{mnSSM,mnSSM1,mnSSM1-1,parameter-space,mnSSM2,mnSSM2-1,Zhang1,Zhang2} can solve the $\mu$ problem~\cite{m-problem} of the minimal supersymmetric standard model (MSSM)~\cite{MSSM,MSSM1,MSSM2,MSSM3,MSSM4}, through introducing three singlet right-handed neutrino superfields $\hat{\nu}_i^c$ ($i=1,2,3$). The neutrino superfields lead the mixing of the neutral components of the Higgs doublets with the right-handed sneutrinos, that is different from the Higgs sector of the MSSM. The mixing can change the Higgs couplings and influence the decay processes of the Higgs bosons. In addition, we also introduce $CP$ violation, and we also get a lighter Higgs at $\sim$ 95 GeV with a suitable parameter space.

The paper is organized as follows. In Sec.~\ref{sec2}, we introduce the $CP$-violating $\mu$-from-$\nu$ SSM briefly, about the superpotential and the CPV phases. In Sec.~\ref{sec3}, we study the excess at 95 GeV in the CPV $\mu$-from-$\nu$ SSM. In Secs.~\ref{sec4} and \ref{sec5}, we show the numerical analysis and the conclusion,respectively.

\section{The Model\label{sec2}}

The superpotential of the $\mu$-from-$\nu$ SSM contains Yukawa couplings for neutrinos, two additional types of terms involving the Higgs doublet superfields $\hat H_u$ and $\hat H_d$, and the right-handed neutrino superfields  $\hat{\nu}_i^c$~\cite{mnSSM}:
\begin{eqnarray}
&&W={\epsilon _{ab}}\left( {Y_{{u_{ij}}}}\hat H_u^b\hat Q_i^a\hat u_j^c + {Y_{{d_{ij}}}}\hat H_d^a\hat Q_i^b\hat d_j^c
+ {Y_{{e_{ij}}}}\hat H_d^a\hat L_i^b\hat e_j^c \right)  \nonumber\\
&&\hspace{0.95cm}
+ {\epsilon _{ab}}{Y_{{\nu _{ij}}}}\hat H_u^b\hat L_i^a\hat \nu _j^c -  {\epsilon _{ab}}{\lambda _i} e^{i \phi_{\lambda_i}} \hat\nu _i^c\hat H_d^a\hat H_u^b + \frac{1}{3}{\kappa _{ijk}}\hat \nu _i^c\hat \nu _j^c\hat \nu _k^c ,
\label{eq-W}
\end{eqnarray}
where $\hat H_u^T = \Big( {\hat H_u^ + ,\hat H_u^0} \Big)$, $\hat H_d^T = \Big( {\hat H_d^0,\hat H_d^ - } \Big)$, $\hat Q_i^T = \Big( {{{\hat u}_i},{{\hat d}_i}} \Big)$, and $\hat L_i^T = \Big( {{{\hat \nu}_i},{{\hat e}_i}} \Big)$ (the index $T$ denotes the transposition) represent the MSSM-like doublet Higgs superfields and $\hat u_i^c$, $\hat d_i^c$, and $\hat e_i^c$ are the singlet up-type quark, down-type quark and charged lepton superfields, respectively.  In addition, $Y_{u,d,e,\nu}$, $\lambda$, and $\kappa$ are dimensionless matrices, a vector, and a totally symmetric tensor, respectively. $a,b=1,2$ are SU(2) indices with antisymmetric tensor $\epsilon_{12}=1$, and $i,j,k=1,2,3$ are generation indices. The $CP$ is violated by the parameter $\lambda_{i}$, and the $CP$-violating phase is $\phi_{\lambda_i}$.

In the superpotential, if the scalar potential is such that nonzero vacuum expectative values (VEVs) of the scalar components ($\tilde \nu _i^c$) of the singlet neutrino superfields $\hat{\nu}_i^c$ are induced, the effective bilinear terms $\epsilon _{ab} \varepsilon_i \hat H_u^b\hat L_i^a$ and $\epsilon _{ab} \mu \hat H_d^a\hat H_u^b$ are generated, with $\varepsilon_i= Y_{\nu _{ij}} \left\langle {\tilde \nu _j^c} \right\rangle$ and $\mu  = {\lambda _i}\left\langle {\tilde \nu _i^c} \right\rangle$,  once the electroweak symmetry is broken. The last term in Eq.~(\ref{eq-W})  generates the effective Majorana masses for neutrinos at the electroweak scale. Therefore, the $\mu$-from-$\nu$ SSM can generate three tiny neutrino masses at the tree level through TeV-scale seesaw mechanism
\cite{mnSSM,neutrino-mass,neu-mass1,neu-mass2,neu-mass3,neu-mass4,neu-mass5,neu-mass6}.

It is worth explaining why the TeV-scale  seesaw was chosen. Through a seesaw on the scale of the grand unified theory, one can get Yukawa couplings of the order of one for neutrinos. But we know that the Yukawa coupling of the electron is on the order of $10^{-6}$, and the Yukawa couplings of neutrinos can also be around on the order of $10^{-6}$ instead of one. In the TeV-scale seesaw, this is sufficient to reproduce the neutrino mass, if the Yukawa coupling of the neutrino is of the same order as the Yukawa coupling of the electron \cite{mnSSM}. Here, it is important to note that the VEVs of the left-handed sneutrinos $\upsilon_{\nu_i}$ are generally small.
We know that the Dirac masses for the neutrinos $m_{D_i}=Y_{\upsilon_{\nu_i}}\upsilon_{u}\lesssim 10^{-4}$ GeV in the TeV-scale seesaw. So we can get an estimate of the VEVs of the left-handed sneutrinos, $\upsilon_{\nu_i}\lesssim m_{D_i}\lesssim 10^{-4}$ GeV, which means that $\upsilon_{\nu_i}\ll\upsilon_{d},\upsilon_{u}$ \cite{mnSSM,mnSSM1}.

The general soft supersymmetry-breaking terms of the $\mu$-from-$\nu$SSM are given by
\begin{eqnarray}
&&- \mathcal{L}_{soft}=m_{{{\tilde Q}_{ij}}}^{\rm{2}}\tilde Q{_i^{a\ast}}\tilde Q_j^a
+ m_{\tilde u_{ij}^c}^{\rm{2}}\tilde u{_i^{c\ast}}\tilde u_j^c + m_{\tilde d_{ij}^c}^2\tilde d{_i^{c\ast}}\tilde d_j^c
+ m_{{{\tilde L}_{ij}}}^2\tilde L_i^{a\ast}\tilde L_j^a  \nonumber\\
&&\hspace{1.7cm} +  m_{\tilde e_{ij}^c}^2\tilde e{_i^{c\ast}}\tilde e_j^c + m_{{H_d}}^{\rm{2}} H_d^{a\ast} H_d^a
+ m_{{H_u}}^2H{_u^{a\ast}}H_u^a + m_{\tilde \nu_{ij}^c}^2\tilde \nu{_i^{c\ast}}\tilde \nu_j^c \nonumber\\
&&\hspace{1.7cm}  +  \epsilon_{ab}{\left[{{({A_u}{Y_u})}_{ij}}H_u^b\tilde Q_i^a\tilde u_j^c
+ {{({A_d}{Y_d})}_{ij}}H_d^a\tilde Q_i^b\tilde d_j^c + {{({A_e}{Y_e})}_{ij}}H_d^a\tilde L_i^b\tilde e_j^c + {\rm{H.c.}} \right]} \nonumber\\
&&\hspace{1.7cm}  + \left[ {\epsilon _{ab}}{{({A_\nu}{Y_\nu})}_{ij}}H_u^b\tilde L_i^a\tilde \nu_j^c
- {\epsilon _{ab}}{{({A_\lambda } e^{i\phi_{\lambda}} \lambda )}_i}\tilde \nu_i^c H_d^a H_u^b
+ \frac{1}{3}{{({A_\kappa }\kappa )}_{ijk}}\tilde \nu_i^c\tilde \nu_j^c\tilde \nu_k^c + {\rm{H.c.}} \right] \nonumber\\
&&\hspace{1.7cm}  -  \frac{1}{2}\left({M_3}{{\tilde \lambda }_3}{{\tilde \lambda }_3}
+ {M_2}{{\tilde \lambda }_2}{{\tilde \lambda }_2} + {M_1}{{\tilde \lambda }_1}{{\tilde \lambda }_1} + {\rm{H.c.}} \right).
\end{eqnarray}
Here, the first two lines contain mass squared terms of squarks, sleptons, and Higgses. The next two lines consist of the trilinear scalar couplings. In the last line, $M_3$, $M_2$, and $M_1$ denote Majorana masses corresponding to SU(3), SU(2), and U(1) gauginos $\tilde{\lambda}_3$, $\tilde{\lambda}_2$, and $\tilde{\lambda}_1$, respectively. In addition to the terms from $\mathcal{L}_{soft}$, the tree-level scalar potential receives the usual $D$- and $F$-term contributions~\cite{mnSSM1,mnSSM1-1}.

Once the electroweak symmetry is spontaneously broken, the neutral scalars develop in general the VEVs. The $CP$ can violated by the VEVs of the scalar fields:
\begin{eqnarray}
\langle H_d^0 \rangle = e^{i\phi_{\upsilon_d}} \upsilon_d , \qquad \langle H_u^0 \rangle = e^{i\phi_{\upsilon_u}} \upsilon_u , \qquad
\langle \tilde \nu_i \rangle = \upsilon_{\nu_i} , \qquad \langle \tilde \nu_i^c \rangle = e^{i\phi_{\upsilon_{\nu_i^c}}} \upsilon_{\nu_i^c} .
\end{eqnarray}
One can define the neutral scalars as
\begin{eqnarray}
&&H_d^0=e^{i\phi_{\upsilon_d}}(\frac{h_d + i P_d}{\sqrt{2}} + \upsilon_d), \qquad\; \tilde \nu_i = \frac{(\tilde \nu_i)^{\text{Re}} + i (\tilde \nu_i)^{\text{Im}}}{\sqrt{2}} + \upsilon_{\nu_i},  \nonumber\\
&&H_u^0=e^{i\phi_{\upsilon_u}}(\frac{h_u + i P_u}{\sqrt{2}} + \upsilon_u), \qquad \tilde \nu_i^c =e^{i\phi_{\upsilon_{\nu_i^c}}}(\frac{(\tilde \nu_i^c)^{\text{Re}} + i (\tilde \nu_i^c)^{\text{Im}}}{\sqrt{2}} + \upsilon_{\nu_i^c}),
\end{eqnarray}
and
\begin{eqnarray}
\tan\beta={\upsilon_u\over{\sqrt{\upsilon_d^2+\upsilon_{\nu_i}\upsilon_{\nu_i}}}}
\approx {\upsilon_u\over\upsilon_d},
\qquad\upsilon=\sqrt{\upsilon_{u}^2+\upsilon_{d}^2+\upsilon_{\nu_i}\upsilon_{\nu_i}}
\approx \sqrt{\upsilon_{u}^2+\upsilon_{d}^2},
\end{eqnarray}

In supersymmetric extensions of the SM, the $R$ parity of a particle is defined as $R = (-1)^{L+3B+2S}$~\cite{MSSM,MSSM1,MSSM2,MSSM3,MSSM4}. $R$ parity is violated if either the baryon number ($B$) or lepton number ($L$) is not conserved, where $S$ denotes the spin of concerned component field. The last two terms in Eq.~(\ref{eq-W}) explicitly violate lepton number and $R$ parity. For example, if one assigns $L = 1$ to the right-handed neutrino superfields, then the last term ${1\over3}\kappa_{ijk}\hat{\nu}_{i}^{c}\hat{\nu}_{j}^{c}\hat{\nu}_{k}^{c}$ in Eq. (2) violates the lepton number by three units contrary to the $\hat{L}_{i}\hat{L}_{j}\hat{e}^{c}_{k}$ term of the $R$ parity violating MSSM which shows the $\Delta L = 1$ effect. $R$ parity breaking implies that the lightest supersymmetric particle is no longer stable.

\subsection{The $\mu$-from-$\nu$SSM Higgs potential\label{subsec1}}

The neutral scalar potential of the tree level can be written as

\begin{eqnarray}
V^{0}=V_{soft}+V_{D}+V_{F}
\end{eqnarray}

with

\begin{eqnarray}
V_{soft}\!\!\!\!\!&&=m_{H_d}^{2}H_{d}^{0}H_{d}^{0\ast}+m_{H_u}^{2}H_{u}^{0}H_{u}^{0\ast}+m_{\tilde{L}_{ij}}^{2}\tilde{\nu}_{i}\tilde{\nu}_{j}^{*}+m_{\tilde{\nu}_{ij}^{c}}^{2} \nonumber \\
&&-((A_{\lambda}e^{i\phi_{\lambda}}\lambda)_{i}\tilde{\nu}_{i}^{c}H_{d}^{0}H_{u}^{0*} -{{1}\over{3}}(A_{\kappa}\kappa)_{ijk}\tilde{\nu}_{i}^{c}\tilde{\nu}_{j}^{c}\tilde{\nu}_{k}^{c}+H.c.)
\end{eqnarray}
\begin{eqnarray}
V_{D}={{G^2}\over{8}}(\tilde{\nu}_{i}\tilde{\nu}_{i}^{*}+H_{d}^{0}H_{d}^{0*}-H_{u}^{0}H_{u}^{0*})^2
\end{eqnarray}
\begin{eqnarray}
V_{F}\!\!\!\!\!&&=\lambda_{i}\lambda_{i}^{*}H_{d}^{0}H_{d}^{0*}H_{u}^{0}H_{u}^{0*}+e^{i\phi_{\lambda_i}}e^{-i\phi_{\lambda_j}}\lambda_{i}\lambda_{j}^{*}e^{-i\phi_{\tilde{\nu}_{i}^{c}}}e^{-i\phi_{\tilde{\nu}_{j}^{c}}}\tilde{\nu}_{i}^{c*}\tilde{\nu}_{j}^{c*} (H_{d}^{0}H_{d}^{0*}+H_{u}^{0}H_{u}^{0*}) \nonumber \\
&&+\kappa_{ijk}\kappa_{ljm}^{*}e^{i\phi_{\tilde{\nu}_{i}^{c}}}e^{i\phi_{\tilde{\nu}_{k}^{c}}}e^{-i\phi_{\tilde{\nu}_{l}^{c}}}e^{-i\phi_{\tilde{\nu}_{m}^{c}}}\tilde{\nu}_{i}^{c}\tilde{\nu}_{k}^{c}\tilde{\nu}_{l}^{c*}\tilde{\nu}_{m}^{c*}
-(e^{-i\phi_{\lambda_{j}}}\lambda_{j}^{*}\kappa_{ijk}\nu_{i}^{c}\nu_{k}^{c}H_{d}^{0}H_{u}^{0*}+h.c.)
\end{eqnarray}
where $G^2=g_{1}^{2}+g_{2}^{2}$. The tree-level neutral scalar potential include the usual soft terms and $D$- and $F$-term contributions. In this work, we take all parameters in the potential are real.

By using the effective potential methods \cite{epm,epm1,epm2,epm3,epm4,epm5,epm6,epm7,epm8,epm9,epm10,epm11,epm12,epm13,epm14,epm15,epm16}, one can get the one-loop effective potential:
\begin{eqnarray}
V^1={{1}\over{32\pi^2}}\left\{\sum_{\tilde{f}}N_{f}m_{\tilde{f}}^{4}\left(\log{{m_{\tilde{f}}^{2}}\over{Q^2}}-{{3}\over{2}}\right)   -2\sum_{f=t,b,\tau}N_{f}m_{f}^{4}\left(\log{{m_{f}^{2}}\over{Q^2}}-{{3}\over{2}}\right)\right\}.
\end{eqnarray}
 Here, $Q$ is the renormalization scale, $N_{t}=N_{b}=3$ and $N_{\tau}=1$. $f=t,b,\tau$ denote the third fermions and $\tilde{f}=\tilde{t}_{1,2}=\tilde{b}_{1,2}=\tilde{\tau}_{1,2}$ are the corresponding supersymmetric partners.

Considering the one-loop effective potential, the Higgs potential can be written as
\begin{eqnarray}
V=V^0+V^1
\end{eqnarray}
we can calculate the minimization conditions of the potential and the Higgs masses in this work.

\subsection{Higgs masses and CPV phases\label{subsec2}}

In the $\mu$-from-$\nu$SSM, the left- and right-handed sneutrino VEVs lead to the mixing of the neutral components of the Higgs doublets with the left- and right-handed sneutrinos producing an $8\times8$ $CP$-even neutral scalar mass matrix, which can be seen in Refs.~\cite{mnSSM1,mnSSM1-1,Zhang1,MASS}. The mixing gives a rich phenomenology in the Higgs sector of the $\mu$-from-$\nu$SSM \cite{mnSSM1,mnSSM2,mnSSM1-1,parameter-space,mnSSM2-1,HZrr,muon,MASS,HZr}.

Here, we note that the Higgs doublets and right-handed sneutrinos are basically decoupled from the left-handed sneutrinos \cite{MASS}, so we did not consider the left-handed sneutrinos in the $CP$-even and $CP$-odd scalars part.

The $CP$-even sector mix with the $CP$-odd sector, the 10 $\times$ 10 mixing matrix is defined by

\begin{eqnarray}
M_{h}^{2}=
\left(
  \begin{array}{cc}
    M_{S}^2 & M_{SP}^{2} \\
    (M_{SP}^{2})^2 & M_{P}^{2} \\
  \end{array}
\right)
\end{eqnarray}
with $M_{S}^{2}$ denoting the $CP$-even neutral scalars, $M_{P}^{2}$ is the $CP$-odd neutral scalars, and $M_{SP}^{2}$ represents the mass submatrix for the mixing of $CP$-even neutral scalars and $CP$-odd neutral scalars.

The mass squared matrix $M_{h}^2$ can be diagonalized as
\begin{eqnarray}
Z_{H}M_{h}^{2}Z_{H}^{T}=m_{h}^2
\end{eqnarray}
with CPV in the $CP$-even and $CP$-odd scalar sector the matrix $Z_{H}$ can be complex.

We consider the radiative corrections in mass submatrix $M_{H}^{2}$; the radiative corrections from the third fermions ($f=t,b,\tau$) and their superpartners include the two-loop leading-log effects \cite{leading-log,leading-log1,leading-log2}. The $CP$-even neutral scalars is given as:
\begin{eqnarray}
M_{S}^{2}=
\left(
  \begin{array}{cc}
    M_{H}^2 & M_{Mix}^{2} \\
    (M_{Mix}^{2})^2 & M_{\tilde{\nu}_{R}^{c}}^{2} \\
  \end{array}
 \right)
\end{eqnarray}

In detail, the mass submatrix $M_{H}^{2}$ is defined by
\begin{eqnarray}
M_{H}^{2}=
\left(
\begin{array}{cc}
M_{h_{d}h_{d}}^{2}+\Delta_{11} & M_{h_{d}h_{u}}^{2}+\Delta_{12} \\
M_{h_{d}h_{u}}^{2}+\Delta_{12} & M_{h_{u}h_{u}}^{2}+\Delta_{22} \\
\end{array}
\right)
\end{eqnarray}

The dominating contributions of radiative corrections $\Delta_{11},~\Delta_{12}$ and $\Delta_{22}$ comes from the fermions ($t, b$) and their superpartners
\begin{eqnarray}
\Delta_{11}=\Delta_{11}^{t}+\Delta_{11}^{b}  \nonumber \\
\Delta_{12}=\Delta_{12}^{t}+\Delta_{12}^{b}  \nonumber \\
\Delta_{22}=\Delta_{22}^{t}+\Delta_{22}^{b}
\end{eqnarray}

We did not consider the terms containing coupling $Y_{\nu_i}$ and $\upsilon_{\nu_{i}}$, because these terms are very small. The radiative corrections from the top quark is given by \cite{HZrr,radiative corr,radiative corr1,radiative corr2,radiative corr3,radiative corr4,radiative corr5}
\begin{eqnarray}
\Delta_{11}^{t}\!\!\!\!\!&&={{3G_{F}m_{t}^{4}}\over{2\sqrt{2}\pi^2\sin^2{\beta}}}{{(\text{Re}(e^{i\phi_{\lambda}}e^{i\phi_{A_{t}}}e^{i\phi_{\upsilon_u}}e^{-i\phi_{\upsilon_{d}}}e^{i\phi_{\upsilon_{\nu_{i}^{c}}}})A_{t}\mu-|\mu|^2\cot\beta)^2    }\over{m_{\tilde{t}_{1}}^{2}-m_{\tilde{t}_{2}}^{2}}}g(m_{\tilde{t}_1}^{2},m_{\tilde{t}_2}^{2})   \\
\Delta_{12}^{t}\!\!\!\!\!&&={{3G_{F}m_{t}^{4}}\over{2\sqrt{2}\pi^2\sin^2{\beta}}}{{|\mu|^2\cot\beta-\text{Re}(e^{i\phi_{\lambda}}e^{i\phi_{A_{t}}}e^{i\phi_{\upsilon_u}}e^{-i\phi_{\upsilon_{d}}}e^{i\phi_{\upsilon_{\nu_{i}^{c}}}})A_{t}\mu   }\over{m_{\tilde{t}_{1}}^{2}-m_{\tilde{t}_{2}}^{2}}} \nonumber \\
&&\times \left(\ln{{m_{\tilde{t}_{1}^2}^{2}}\over{m_{\tilde{t}_{2}}^{2}}}+{{\text{Re}(e^{i\phi_{\lambda}}e^{i\phi_{A_{t}}}e^{i\phi_{\upsilon_u}}e^{-i\phi_{\upsilon_{d}}}e^{i\phi_{\upsilon_{\nu_{i}^{c}}}})A_{t}\mu\cot\beta  }\over{m_{\tilde{t}_{1}}^{2}-m_{\tilde{t}_{2}}^{2}}}g(m_{\tilde{t}_1}^{2},m_{\tilde{t}_2}^{2}) \right) \\
\Delta_{22}^{t}\!\!\!\!\!&&={{3G_{F}m_{t}^{4}}\over{2\sqrt{2}\pi^2\sin^2{\beta}}}\Bigg\{\ln\frac{m_{\tilde{t}_1}^{2}m_{\tilde{t}_2}^{2}}{m_{t^{4}}}+{{2|A_{t}|^2-\text{Re}(e^{i\phi_{\lambda}}e^{i\phi_{A_{t}}}e^{i\phi_{\upsilon_u}}e^{-i\phi_{\upsilon_{d}}}e^{i\phi_{\upsilon_{\nu_{i}^{c}}}})A_{t}\mu\cos\beta   }\over{m_{\tilde{t}_{1}}^{2}-m_{\tilde{t}_{2}}^{2}}}\ln{{m_{\tilde{t}_{1}^2}^{2}}\over{m_{\tilde{t}_{2}}^{2}}} \nonumber \\
&&+{{(2|A_{t}|^2-\text{Re}(e^{i\phi_{\lambda}}e^{i\phi_{A_{t}}}e^{i\phi_{\upsilon_u}}e^{-i\phi_{\upsilon_{d}}}e^{i\phi_{\upsilon_{\nu_{i}^{c}}}})A_{t}\mu\cot\beta)^2}\over{(m_{\tilde{t}_{1}}^{2}-m_{\tilde{t}_{2}}^{2})^2}}g(m_{\tilde{t}_1}^{2},m_{\tilde{t}_2}^{2})
+{{1}\over{16\pi^2}}\ln{{m_{\tilde{t}_{1}}^{2}m_{\tilde{t}_{2}}^{2}}\over{m_{t}^{4}}}  \nonumber\\
&&\times \left({{3e^2m_{t}^2}\over{4s_{w}^2m_{W}^2}}-32\pi\alpha_{s}\right)\Bigg[{1\over2}\ln{{m_{\tilde{t}_{1}}^{2}m_{\tilde{t}_{2}}^{2}}\over{m_{t}^4}}+{{2|A_{t}-\mu\cot\beta|^2}\over{m_{\tilde{t}_{1}}m_{\tilde{t}_{2}}}}  \Bigg(1-{{(A_{t}-\mu\cot\beta)^2}\over{12m_{\tilde{t}_{1}}^{2}m_{\tilde{t}_{2}}^{2}}} \Bigg)\Bigg]    \Bigg\}
\end{eqnarray}
with
\begin{eqnarray}
g(m_{1}^{2},m_{2}^{2})=2-{{m_{1}^{2}+m_{2}^{2}}\over{m_{1}^{2}-m_{2}^{2}}}\ln{{m_{1}^{2}}\over{m_{2}^{2}}}
\end{eqnarray}
to save space, the mass matrix and the radiative corrections are given in the Appendix.

In the radiative corrections, the trilinear coupling $A_{t}=|A_{t}| e^{i\phi_{A_{t}}}$ can be complex. These seven independent phase have been defined as
\begin{eqnarray}
\phi_{\lambda}, \phi_{A_{t}}, \phi_{\upsilon_u}, \phi_{\upsilon_{d}}, \phi_{\upsilon_{\nu_{i}^{c}}}.
\end{eqnarray}

\section{Excess at 95 GeV  \label{sec3}}

The process measured at LEP reported a 2.3$\sigma$ local excess in the $b\bar{b}$ final state searches, with the scalar mass at $\sim$ 96 GeV. The production of a Higgs boson via Higgstrahlung is associated with the Higgs decaying to bottom quarks. Normalized to the SM expectation, the signal strength is defined as
\begin{eqnarray}
\mu_{LEP}^{bb}={{\sigma(e^{+}e^{-}\rightarrow Zh_{1}\rightarrow Zb\bar{b})}\over{\sigma^{SM}(e^{+}e^{-}\rightarrow Zh_{1}\rightarrow Zb\bar{b})}}=0.117\pm 0.057
\end{eqnarray}

The value for $\mu_{LEP}^{bb}$ can be found in \cite{98LEP1,98LHC} with the method introduced in \cite{98LEP2}. $h_1$ is the Higgs which has mass range around $\sim$ 96 GeV, and $h_2$ is the 125 GeV Higgs boson in the following. In the framework of the $\mu$-from-$\nu$ SSM we use $\mu_{NP}^{bb}$ to describe the signal strength; the expression for $\mu_{NP}^{bb}$ can be approximated as \cite{98LEP3}
\begin{eqnarray}
\mu_{NP}^{bb}\!\!\!\!\!&&={{\sigma^{\text{NP}}(Z^*\rightarrow Zh_{1})}\over{\sigma^{\text{SM}}(Z^*\rightarrow Zh_{1})}}\times {{Br^{\text{NP}}(h_{1}\rightarrow b\bar{b})}\over{Br^{\text{SM}}(h_{1}\rightarrow b\bar{b})}} \nonumber \\
&&\approx |C_{h_{1}VV}|^2\times {{\Gamma^{\text{NP}}_{h_1\rightarrow b\bar{b}}}\over{\Gamma_{h_1\rightarrow b\bar{b}}^{\text{SM}}}}\times {{\Gamma_{\text{tot}}^{\text{SM}}}\over{\Gamma_{\text{tot}}^{\text{NP}}}} \nonumber \\
&&\approx {{|C_{h_{1}VV}|^2\times |C_{h_{1}d\bar{d}}|^2}\over{|C_{h_{1}d\bar{d}}|^2(Br_{h_1\rightarrow b\bar{b}}^{\text{SM}}+Br_{h_1\rightarrow \tau\bar{\tau}}^{\text{SM}})+|C_{h_{1}u\bar{u}}|^2(Br_{h_1\rightarrow g\bar{g}}^{\text{SM}}+Br_{h_1\rightarrow c\bar{c}}^{\text{SM}})}}
\label{muLEP}
\end{eqnarray}

One can find the SM branching ratios $Br^{SM}$ in Ref. \cite{PDG1}, and $\Gamma$ is the decay widths. The couplings are normalized to the SM prediction of a Higgs boson of the same mass. $C_{h_{1}}$ is the coupling of $h_1$ and gauge boson, and $C_{h_1u\bar{u}}$ and $C_{h_1d\bar{d}}$ are the couplings of $h_1$ and up- and down-type quarks. The normalized couplings are given as
\begin{eqnarray}
C_{h_{i}d\bar{d}}={{Z_{H}^{i1}}\over{\cos\beta}}, \qquad C_{h_{i}u\bar{u}}={{Z_{H}^{i2}}\over{\sin\beta}}, \qquad C_{h_{i}VV}=Z_{H}^{i1}\cos\beta+Z_{H}^{i2}\sin\beta.
\end{eqnarray}

In 2019, the CMS searches for the Higgs boson decaying in the diphoton channel showed a local excess of $\sim$ 3$\sigma$ around $\sim$ 96 GeV \cite{CMS3}; the previous results is that \cite{CMS3,98CMS}
\begin{eqnarray}
\mu_{\gamma\gamma}^{CMS}={{ \sigma(gg\rightarrow h_{1}\rightarrow\gamma\gamma) }\over{ \sigma^{SM}(gg\rightarrow h_{1}\rightarrow\gamma\gamma) }}=0.6\pm 0.2.
\end{eqnarray}

Recently, ATLAS reported their new results at 95.4 GeV based on the full Run 2 dataset \cite{95ATLAS}, the ``diphoton excess'' with a signal strength of
\begin{eqnarray}
\mu_{\gamma\gamma}^{ATLAS}=0.18\pm0.1
\end{eqnarray}
Meanwhile, the corresponding CMS result for the ``diphoton excess'' is given by \cite{CMS4}
\begin{eqnarray}
\mu_{\gamma\gamma}^{CMS}={{\sigma^{exp}(pp\rightarrow \phi\rightarrow\gamma\gamma)}\over{\sigma^{SM}(pp\rightarrow H_{SM}\rightarrow\gamma\gamma)}}=0.33_{-0.12}^{+0.19}.
\end{eqnarray}
Neglecting possible correlations one can get a combined signal strength of \cite{CMS5}
\begin{eqnarray}
\mu_{\gamma\gamma}^{exp}=\mu_{\gamma\gamma}^{ATLAS+CMS}=0.24_{-0.08}^{+0.09}.
\end{eqnarray}

In this work, the approximation of the diphoton rate of the $h_{1}$ can written as \cite{98LEP1,98LEP3}
\begin{eqnarray}
\mu_{NP}^{\gamma\gamma}\!\!\!\!\!&&={{\sigma^{\text{NP}}(gg\rightarrow h_{1})}\over{\sigma^{\text{SM}}(gg\rightarrow h_{1})}}\times {{Br^{\text{NP}}(h_1\rightarrow \gamma\gamma)}\over{Br^{\text{SM}}(h_1\rightarrow \gamma\gamma)}} \nonumber \\
&&\approx |C_{h_{1}u\bar{u}}|^2\times {{\Gamma_{h_1\rightarrow\gamma\gamma}^{\text{NP}}}\over{\Gamma_{h_1\rightarrow\gamma\gamma}^{\text{SM}}}}\times {{\Gamma_{\text{tot}}^{\text{SM}}}\over{\Gamma_{\text{tot}}^{\text{NP}}}} \nonumber \\
&&\approx {{|C_{h_{1}u\bar{u}}|^2\times|C_{h_{1}\gamma\gamma}|^2}\over{|C_{h_{1}d\bar{d}}|^2(Br_{h_1\rightarrow b\bar{b}}^{\text{SM}}+Br_{h_1\rightarrow \tau\bar{\tau}}^{\text{SM}})+|C_{h_{1}u\bar{u}}|^2(Br_{h_1\rightarrow gg}^{\text{SM}}+Br_{h_1\rightarrow c\bar{c}}^{\text{SM}})  }}.
\label{muCMS}
\end{eqnarray}

The effective coupling $C_{h_1 \gamma\gamma}$ can be written as \cite{98LEP3}
\begin{eqnarray}
|C_{h_{1}\gamma\gamma}|^2={{|{4\over3}C_{h_1t\bar{t}}A_{1/2}(\tau_{t})+C_{h_1VV}A_{1}(\tau_W)|^2}\over{|{4\over3}A_{1/2}(\tau_t)+A_{1}(\tau_{W})|^2 }}
\end{eqnarray}
with $\tau_{t}={{m_{h_1}^{2}}\over{4m_{t}^{2}}}<1$ and $\tau_{t}={{m_{h_1}^{2}}\over{4m_{W}^{2}}}<1$. The form factors $A_{1/2}$ and $A_{1}$ are given by \cite{Chrr}
\begin{eqnarray}
A_{1/2}(x)\!\!\!\!\!&&=2(x+(x-1)\arcsin^2\sqrt{x})x^{-2},  \qquad\qquad\qquad x\leq1 \\
A_{1}(x)\!\!\!\!\!&&=-(2x^2+3x+3(2x-1)\arcsin^2\sqrt{x})x^{-2} \qquad x\leq1
\end{eqnarray}
By using Eqs. (\ref{muLEP}) and (\ref{muCMS}), we calculate the two signal strengths.
\section{Numerical Results\label{sec4}}
In this section we will discuss the couplings and signal strength of 96 GeV Higgs in CPV $\mu\nu$SSM. The free parameters in our analysis will be
\begin{eqnarray}
\lambda,\quad \tan\beta,\quad \kappa,\quad A_{t},\quad A_{\lambda},\quad A_{\kappa},\quad A_{b},\quad \upsilon_{\nu^{c}}.
\end{eqnarray}
We take $\upsilon_{\nu_{1}^{c}}=\upsilon_{\nu_{2}^{c}}=\upsilon_{\nu_{3}^{c}}$, and we have defined
\begin{eqnarray}
&&\hspace{-0.9cm}{\kappa _{ijk}} = \kappa {\delta _{ij}}{\delta _{jk}}, \quad
{({A_\kappa }\kappa )_{ijk}} =
{A_\kappa }\kappa {\delta _{ij}}{\delta _{jk}}, \quad
\lambda _i = \lambda , \nonumber\\
&&\hspace{-0.9cm}
{({A_\lambda }\lambda )}_i = {A_\lambda }\lambda,\quad
{Y_{{e_{ij}}}} = {Y_{{e_i}}}{\delta _{ij}},\quad
{({A_e}{Y_e})_{ij}} = {A_{e}}{Y_{{e_i}}}{\delta _{ij}},\nonumber\\
&&\hspace{-0.9cm}
{Y_{{\nu _{ij}}}} = {Y_{{\nu _i}}}{\delta _{ij}},\quad
(A_\nu Y_\nu)_{ij}={a_{{\nu_i}}}{\delta _{ij}},\quad
m_{\tilde \nu_{ij}^c}^2 = m_{\tilde \nu_{i}^c}^2{\delta _{ij}}, \nonumber\\
&&\hspace{-0.9cm}m_{\tilde Q_{ij}}^2 = m_{{{\tilde Q_i}}}^2{\delta _{ij}}, \quad
m_{\tilde u_{ij}^c}^2 = m_{{{\tilde u_i}^c}}^2{\delta _{ij}}, \quad
m_{\tilde d_{ij}^c}^2 = m_{{{\tilde d_i}^c}}^2{\delta _{ij}}, \nonumber\\
&&\hspace{-0.9cm}m_{{{\tilde L}_{ij}}}^2 = m_{{\tilde L}}^2{\delta _{ij}}, \quad
m_{\tilde e_{ij}^c}^2 = m_{{{\tilde e}^c}}^2{\delta _{ij}}, \quad
\upsilon_{\nu_i^c}=\upsilon_{\nu^c},
\label{MFV}
\end{eqnarray}
where $i,j=1,2,3$.

\subsection{Mass and coupling}
\begin{figure}
\setlength{\unitlength}{1mm}
\begin{minipage}[c]{0.5\textwidth}
\centering
\includegraphics[width=3in]{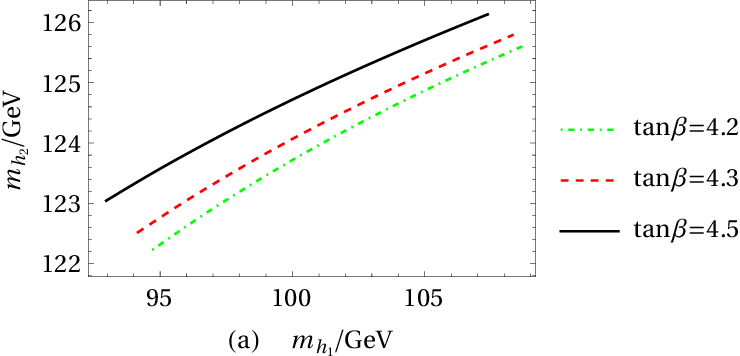}
\end{minipage}%
\begin{minipage}[c]{0.5\textwidth}
\centering
\includegraphics[width=3in]{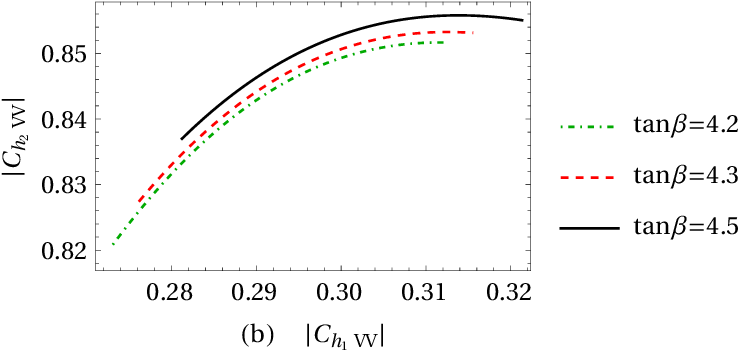}
\end{minipage}%
\caption[]{Values for $m_{h_2}$ versus $m_{h_2}$ in (a) and the normalized couplings $|C_{h_{2}VV}|$ versus $|C_{h_{1}VV}|$ in (b). The input parameters are in Table \ref{para}.}
\label{mass-hvv}
\end{figure}
In Fig. \ref{mass-hvv} (a), we can see that, when $h_{1}$ is near 96 GeV, $h_{2}$ is closer to 125 GeV with the increase of $\tan\beta$. Although both $h_{1}$ and $h_{2}$ can conform to the experimental mass range in our parameter space, if we assume a theory uncertainty of up to 3 GeV \cite{98LEP3}, the parameter range will be larger. One can clear see that $|C_{h_{2}VV}|$ is much larger than $|C_{h_{1}VV}|$ in Fig. \ref{mass-hvv} (b); the LHC measurements of the SM-like Higgs bosons couplings to fermions and massive gauge bosons are still not very precise\cite{98LEP3,LHC-couplings}. If in the future some collider can measure these couplings to the percent level, then we can choose a more reasonable parameter space.

\begin{figure}
\setlength{\unitlength}{1mm}
\begin{minipage}[c]{0.85\textwidth}
\centering
\includegraphics[width=5.5in]{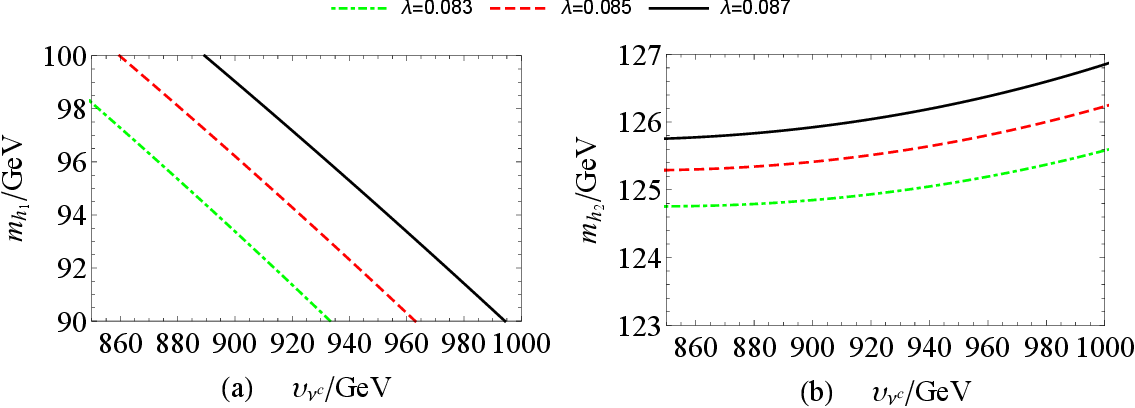}
\end{minipage}%
\\
\begin{minipage}[c]{0.85\textwidth}
\centering
\includegraphics[width=5.5in]{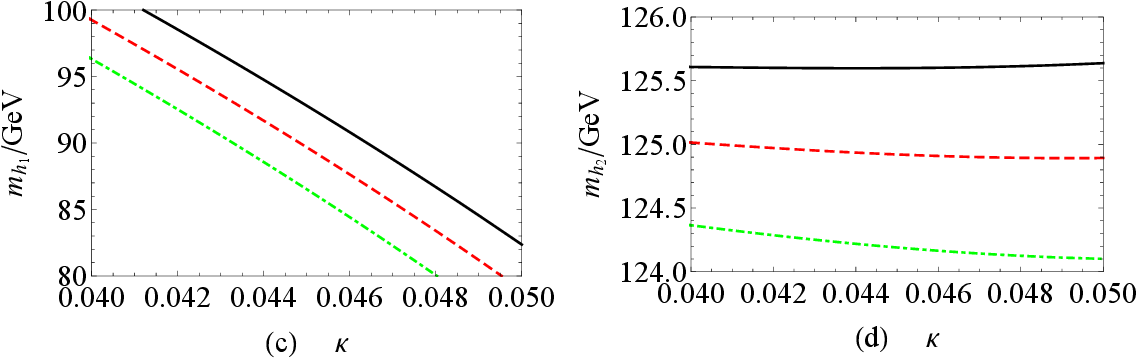}
\end{minipage}%
\caption[]{Values for $m_{h_1}$ versus the parameters $\upsilon_{\nu^c}$ in (a) and $\kappa$ in (c) and values for $m_{h_2}$ versus the parameters $\upsilon_{\nu^c}$ in (b) and $\kappa$ in (d). We take $\tan\beta=5$ and other parameters in Table \ref{para}.}
\label{vc-k}
\end{figure}
In Fig.\ref{vc-k}, left column, we can see that $m_{h_1}$ is very sensitive to $\upsilon_{\nu^c}$. As $\upsilon_{\nu^c}$ becomes larger, $m_{h_1}$ will rapidly become smaller; on the contrary, $m_{h_2}$ will slowly become larger. In order for $m_{h_1}$ to be around 95.4 GeV, the value of $\upsilon_{\nu^c}$ will not have a wide range. Similar to $\upsilon_{\nu^c}$, $m_{h_1}$ is also very sensitive to $\kappa$, and with the increase of $\lambda$, a larger $\kappa$ value can be taken to keep $m_{h_1}$ near 95.4 GeV. On the contrary, $\kappa$ has little effect on $m_{h_2}$; especially when $\lambda$=0.085 or 0.087, $m_{h_2}$ will be very stable.

\begin{table}
\scriptsize
\begin{tabular}{|c|c|c|c|c|c|}
  \hline
  $\qquad A_{\lambda} \qquad$  & $\qquad \lambda \qquad$  &  $\qquad A_t \qquad$  & $\qquad \upsilon_{\nu^{c}} \qquad$ & $\qquad \kappa \qquad$ & $\qquad A_{\kappa}\qquad$ \\ \hline
  920 & [0.08;0.09] & 1660 & 1000 & 0.042 & $-$370 \\ \hline
  $\qquad A_{b} \qquad$ & $\qquad \phi_{\lambda}/\pi \qquad$ & $\qquad \phi_{A_{t}}/\pi \qquad$ & $\qquad \phi_{v_{u}}/\pi \qquad$ & $\qquad \phi_{v_{d}}/\pi \qquad$ & $\qquad \phi_{\upsilon_{\nu^{c}}}/\pi \qquad$ \\  \hline
  700 & $-0.020$ & $-$0.200 & 0.380 & $-$0.095 & $-$1 \\
  \hline
\end{tabular}
\caption{Input parameters to fit the LEP and the CMS excesses. All dimensionful parameters are given in GeV.}
\label{para}
\end{table}
\begin{figure}
\setlength{\unitlength}{1mm}
\begin{minipage}[c]{0.9\textwidth}
\centering
\includegraphics[width=5.5in]{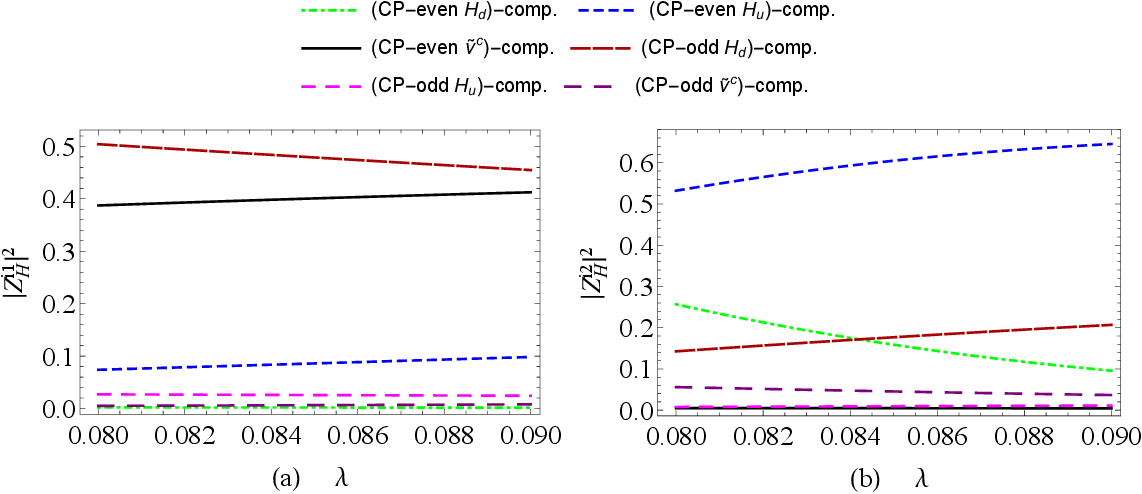}
\end{minipage}%
\caption[]{The component of $h_1$ (a) and $h_2$ (b). The input parameters are in Table \ref{para}, and we take $\tan\beta=4.31$. }
\label{higgs}
\end{figure}

In Fig. \ref{higgs}, we have showed the component of $h_{1}$ and $h_{2}$; for $h_1$, $CP$-odd $H_{u}$ component and $CP$-even right-handed sneutrinos component are the main components, and with the increase of $\lambda$, $CP$-odd $H_{d}$ component will become smaller, while the $CP$-even right-handed sneutrino component will become larger. While for $h_2$, $CP$-even $H_{u}$ is the main component, as $\lambda$ increases, the $CP$-even $H_{u}$ component will increase, the $CP$-even $H_{d}$ component will gradually decrease, and the $CP$-odd $H_d$ component will gradually increase.

\begin{figure}
\setlength{\unitlength}{1mm}
\begin{minipage}[c]{0.5\textwidth}
\centering
\includegraphics[width=2.6in]{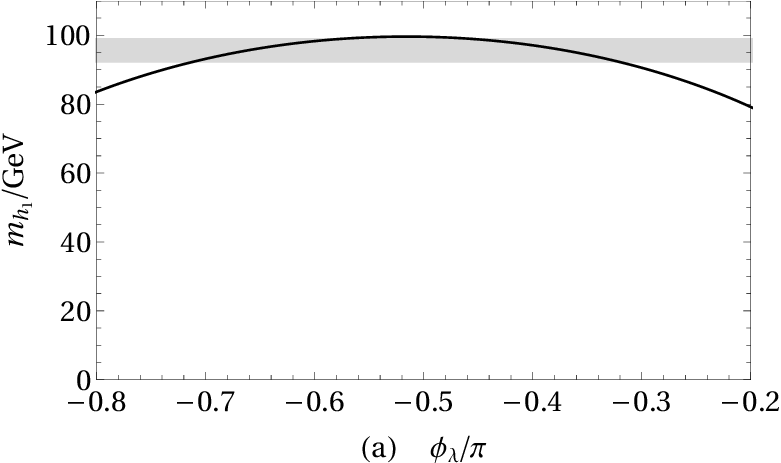}
\end{minipage}%
\begin{minipage}[c]{0.5\textwidth}
\centering
\includegraphics[width=2.6in]{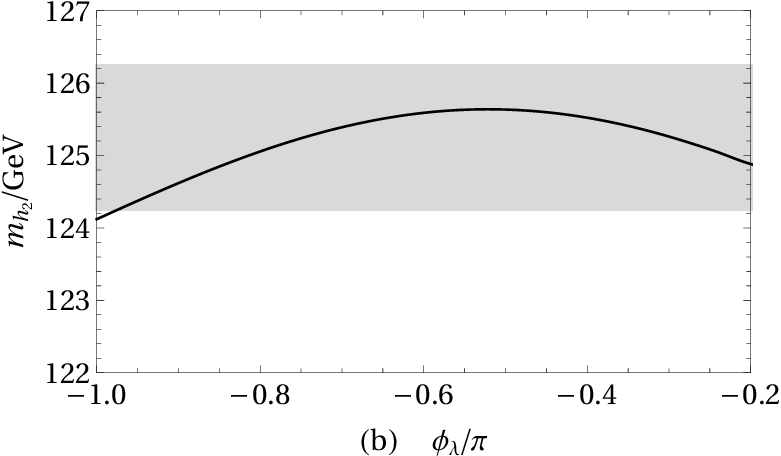}
\end{minipage}%
\\
\begin{minipage}[c]{0.5\textwidth}
\centering
\includegraphics[width=2.6in]{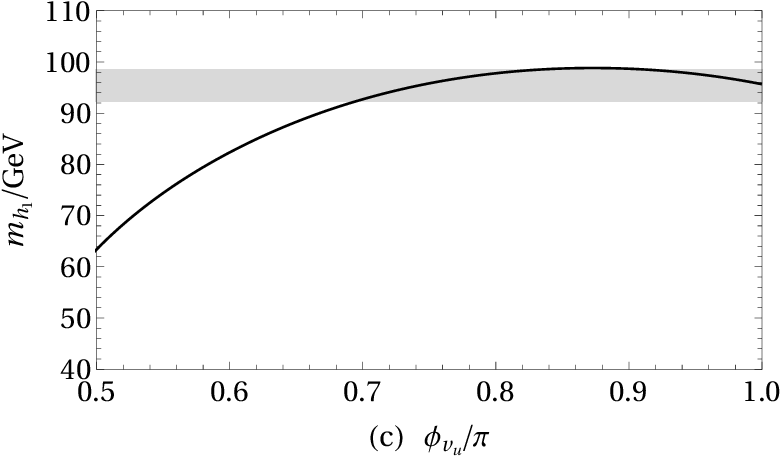}
\end{minipage}%
\begin{minipage}[c]{0.5\textwidth}
\centering
\includegraphics[width=2.6in]{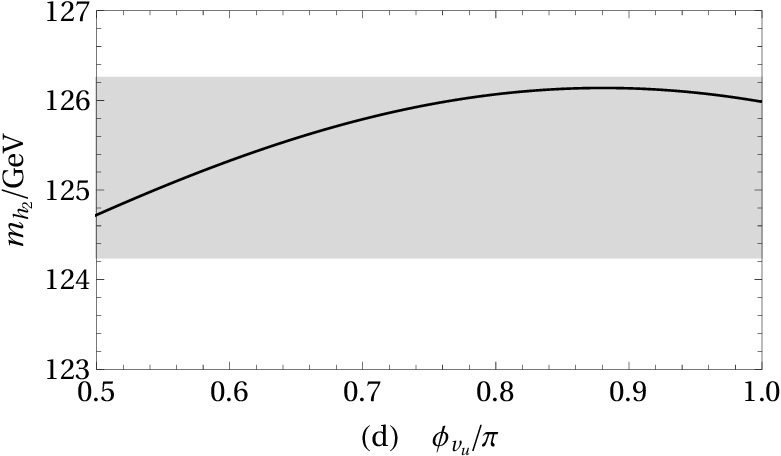}
\end{minipage}%
\\
\begin{minipage}[c]{0.5\textwidth}
\centering
\includegraphics[width=2.6in]{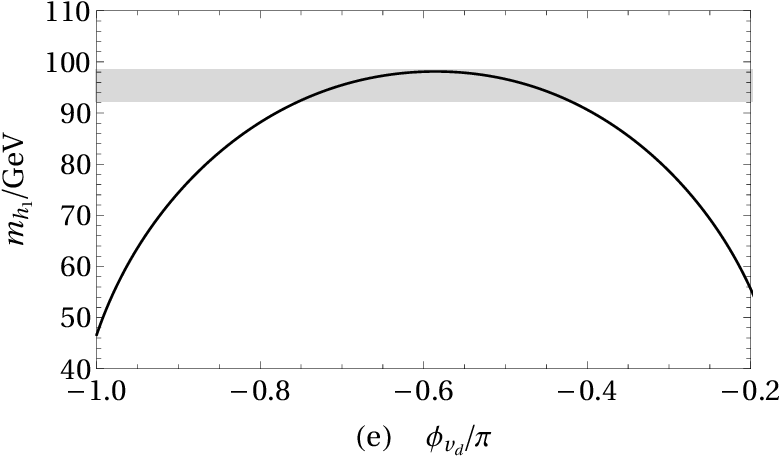}
\end{minipage}%
\begin{minipage}[c]{0.5\textwidth}
\centering
\includegraphics[width=2.6in]{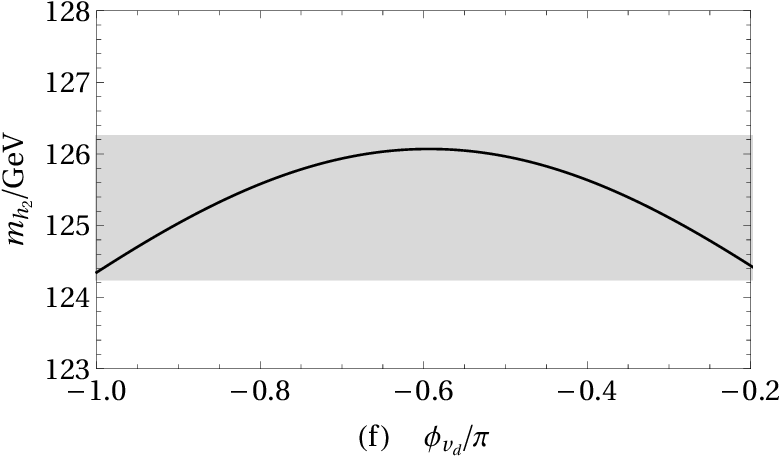}
\end{minipage}%
\\
\begin{minipage}[c]{0.5\textwidth}
\centering
\includegraphics[width=2.6in]{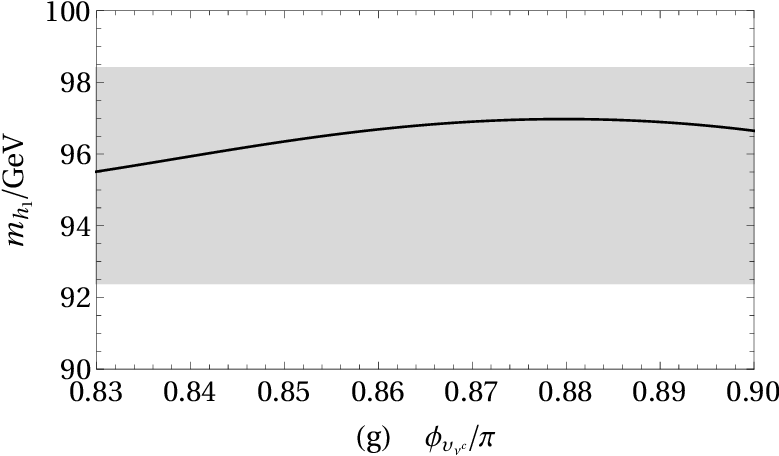}
\end{minipage}%
\begin{minipage}[c]{0.5\textwidth}
\centering
\includegraphics[width=2.6in]{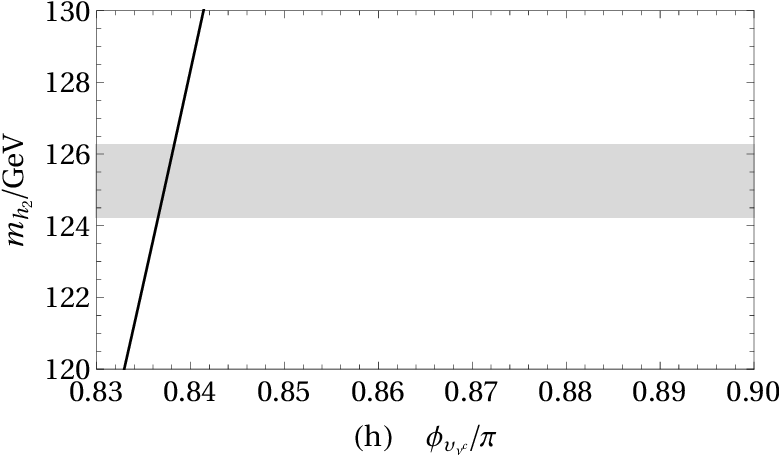}
\end{minipage}%
\caption[]{The correlation of $h_1$ and the CPV phases in (a) (c) (e) (g); the gray area represents the experimental error of 3 GeV. The correlation of $h_2$ and the CPV phases in (b) (d) (f) (h); the gray area is the theory uncertainty of 1 GeV. }
\label{SS-phi}
\end{figure}
Then, in Fig. \ref{SS-phi}, we analyze the correlation between the Higgs masses and the CPV phases. For the first row, we take $\lambda=0.09, \tan\beta=6,$ and $\kappa=0.315,$ for the second row, $\lambda=0.095, \tan\beta=4.5,$ and $\kappa=0.315,$ for the third row, $\lambda=0.098, \tan\beta=4.3,$ and $\kappa=0.32$, for the last row, $\lambda=0.085, \tan\beta=3,$ and $\kappa=0.04$. Other parameters remain the same as Table \ref{para}. We should remark that there are many CPV phase values that can be constrained by the Higgs masses in Fig.\ref{gaga-bb} (a), and here we select only one of them. The signal strengths and the Higgs masses do not increase or decrease all the time, but oscillate like a sine function. In Fig. \ref{SS-phi}, the first row, when $\phi_{\lambda}$ grows from 4 to 5, the mass range of the ``diphoton excess'' can fit the experimental constraints, and the SM-like Higgs mass can be kept around 125 GeV.

Similar to the first row in Fig. \ref{SS-phi}, the second and third rows show the variation of $h_1$ and $h_2$ with the phases $\phi_{v_u}$ and $\phi_{v_d}$. $m_{h_1}$ is very sensitive to $\phi_{v_u}$ and $\phi_{v_d}$. When $\phi_{v_u}$ is 0.7$\pi-\pi$, $h_1$ can be kept within the experimental error of 3 GeV. When $\phi_{v_u}$ continues to increase or decrease, $m_{h_1}$ will drop very quickly. Similarly, when $\phi_{v_d}$ is $[-0.75\pi, -0.45\pi]$, $m_{h_1}$ can conform to the experimental error. For SM-like Higgs, $\phi_{v_u}$ and $\phi_{v_d}$ can provide a wide range to keep $m_{h_2}$ near 125 GeV. Different from other CPV phases, it is difficult to determine the value of $\phi_{\upsilon_{\nu}^c}$, and only a narrow range allows $h_1$ and $h_2$ to be simultaneously constrained by their respective experiments.

\subsection{Signal strengths}
\begin{figure}
\setlength{\unitlength}{1mm}
\begin{minipage}[c]{0.85\textwidth}
\centering
\includegraphics[width=5.5in]{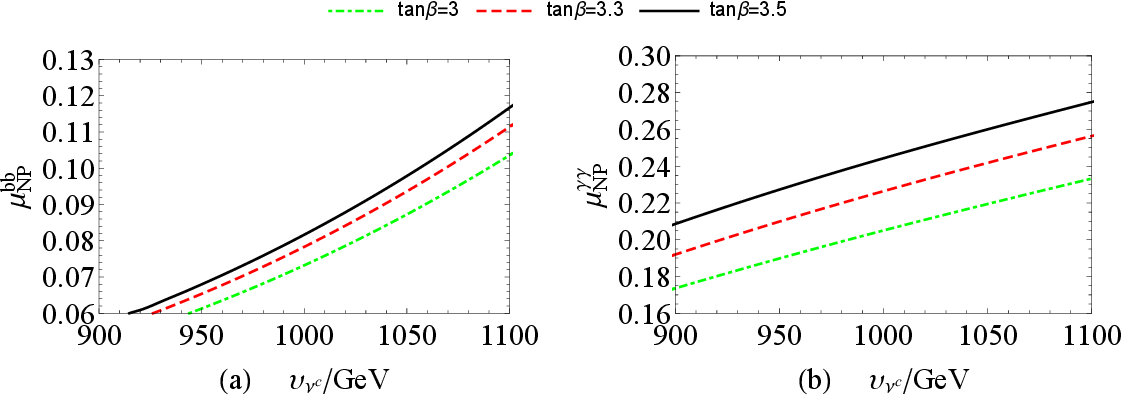}
\end{minipage}%
\\
\begin{minipage}[c]{0.85\textwidth}
\centering
\includegraphics[width=5.5in]{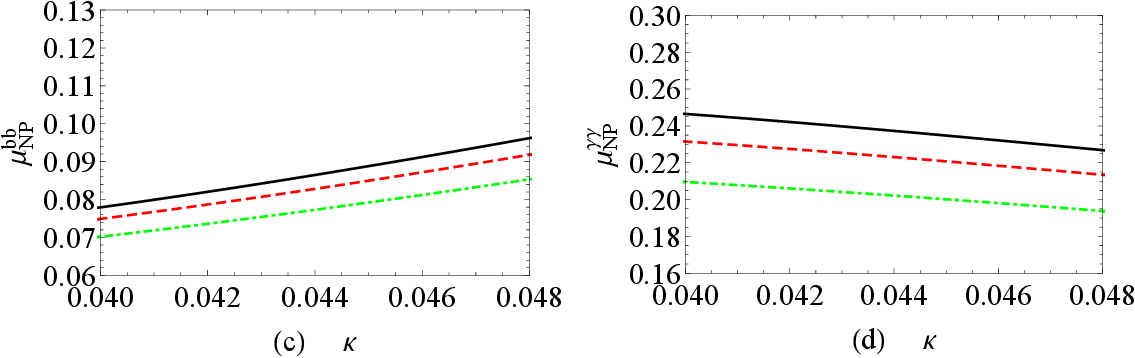}
\end{minipage}%
\caption[]{The left column shows that the signal strength $\mu_{NP}^{bb}$ varies with $\upsilon_{\nu^c}$ (a) and $\kappa$ (c); the right column shows that the signal strength $\mu_{NP}^{\gamma\gamma}$ varies with $\upsilon_{\nu^c}$ (b) and $\kappa$ (d). We take $\lambda=0.09$ and other parameters as in Table \ref{para}.}
\label{mu-vc-k}
\end{figure}
First, in Figs.\ref{mu-vc-k} (a) and (b), we can see the correlation of the signal strengths $\mu_{NP}^{bb}$, $\mu_{NP}^{\gamma\gamma}$ and $\upsilon_{\nu^c}$; with the increase of the value of $\upsilon_{\nu^c}$, the values of $\mu_{NP}^{bb}$ and $\mu_{NP}^{\gamma\gamma}$ will increase. And if we set $\upsilon_{\nu^c}=1000$, the value of $\mu_{NP}^{bb}$ of $\tan\beta$=3.5 will be larger than that of $\tan\beta$=3, because we conclude from Fig. \ref{mass-hvv} that the Higgs masses and couplings will increase with the increase of $\tan\beta$. In the second row in Fig.\ref{mu-vc-k}, as the growth of $\kappa$, the value of $\mu_{NP}^{bb}$ will slowly increase, while the value of $\mu_{NP}^{\gamma\gamma}$ will slowly decrease. However, either $\tan\beta=3$ or $\tan\beta=3.5$ ensures that $\mu_{NP}^{bb}$ and $\mu_{NP}^{\gamma\gamma}$ are both within their respective 1$\sigma$ experimental error.

\begin{figure}
\setlength{\unitlength}{1mm}
\begin{minipage}[c]{0.5\textwidth}
\centering
\includegraphics[width=3in]{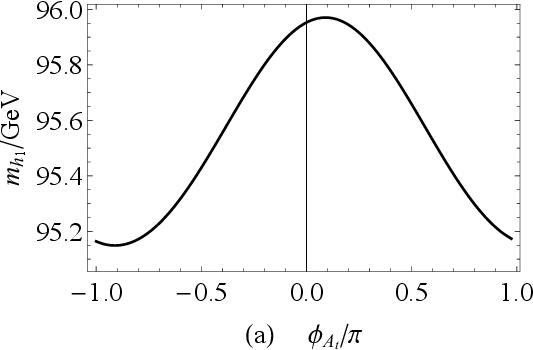}
\end{minipage}%
\begin{minipage}[c]{0.5\textwidth}
\centering
\includegraphics[width=3in]{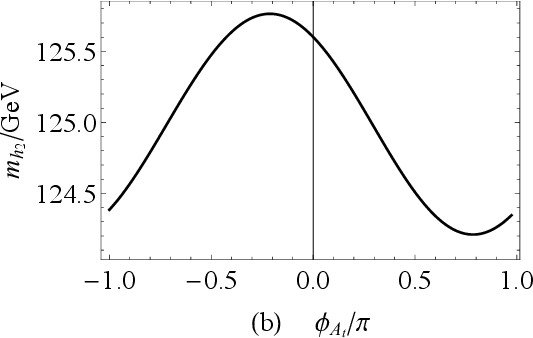}
\end{minipage}%
\\
\begin{minipage}[c]{0.5\textwidth}
\centering
\includegraphics[width=3in]{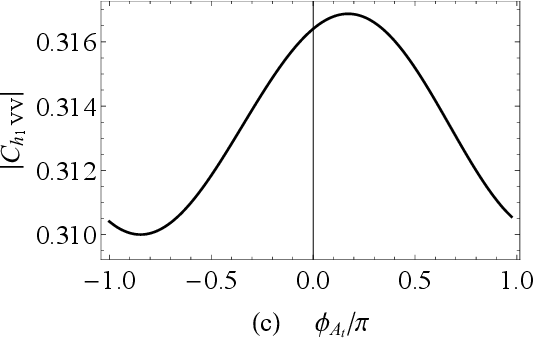}
\end{minipage}%
\begin{minipage}[c]{0.5\textwidth}
\centering
\includegraphics[width=3in]{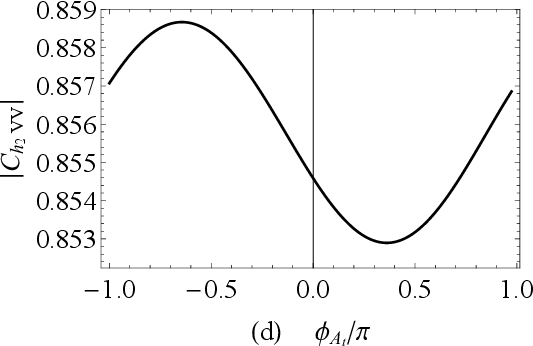}
\end{minipage}%
\\
\begin{minipage}[c]{0.5\textwidth}
\centering
\includegraphics[width=3in]{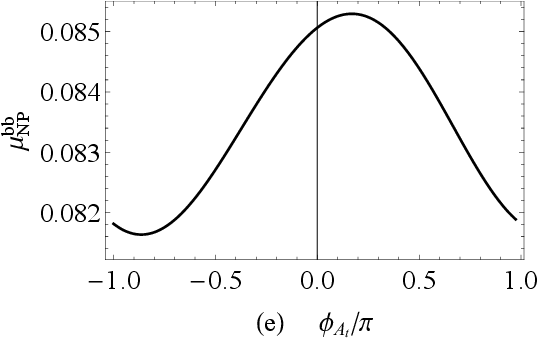}
\end{minipage}%
\begin{minipage}[c]{0.5\textwidth}
\centering
\includegraphics[width=3in]{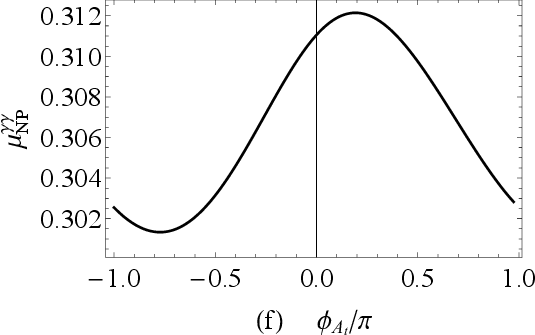}
\end{minipage}%
\caption[]{The Higgs masses vary with the phase $\phi_{A_t}$ in (a)-(b); the correlation of couplings and the phase $\phi_{A_t}$ in (c)-(d); and (e)-(f) are the correlation of signal strengths and the phase $\phi_{A_t}$. Here, we take $\tan\beta$=4.5, $\lambda$=0.086, and $\phi_{A_t}=[-\pi,\pi]$, and other parameters are the same as in Table \ref{para}.}
\label{Phi-mass}
\end{figure}

The effect of CPV phases is also very obvious. Here, we take $\phi_{A_t}$ as an example, we take $\phi_{A_t}$ from $-\pi$ to $\pi$. In the first row in Fig. \ref{Phi-mass}, we can see that the minimum value of $h_{1}$ appears to be around 96.2 GeV and varies periodically with $\phi_{A_t}$. The maximum value of $h_{1}$ does not exceed 1 GeV larger than the minimum value. Meanwhile, the highest point of $h_{2}$ is at 125.3 GeV, with a maximum value 1.5GeV larger than the minimum value.

The impact of $\phi_{A_t}$ on the normalized couplings is relatively small. In the second row in Fig. \ref{Phi-mass}, we can see that the peak of $|C_{h_1VV}|$ is close to 0.313, and the highest point of $|C_{h_2VV}|$ is slightly less than 0.856. This implies that, if the "95.4 excess" is a new particle, then its coupling with gauge bosons should be much smaller than the coupling of the SM-like Higgs with gauge bosons.

In the last row in Fig. \ref{Phi-mass}, we found that $\phi_{A_t}$ also has a slight effect on the two signal strengths, $\mu_{NP}^{\gamma\gamma}$ is always slightly bigger than the central value of $\mu_{exp}^{\gamma\gamma}$, but does not exceed 1$\sigma$ experimental error. At the same time, $\mu_{NP}^{bb}$ is lower than the central value of $\mu_{exp}^{bb}$, but it also does not exceed 1 $\sigma$ experimental error.
\begin{table}
\scriptsize
\begin{tabular}{|c|c|c|c|c|c|c|}
  \hline
  $\qquad \lambda \qquad$  & $\qquad \tan\beta \qquad$  &  $\qquad \kappa \qquad$ &  $\qquad \upsilon_{\nu^{c}} \qquad$ &  $\qquad A_{\lambda} \qquad$ &  $\qquad A_{t} \qquad$ &  $\qquad A_{\kappa} \qquad$ \\ \hline
  $[0.087;0.090]$ & [2.4;4.5] & [0.040;0.045] & [950;1050] & [915;920] & [1600;1700] & [$-$375;$-$370] \\ \hline
  $\qquad A_{b} \qquad$ &  $\qquad \phi_{\lambda}/\pi \qquad$ &  $\qquad \phi_{A_{t}}/\pi \qquad$ &  $\qquad \phi_{v_{u}}/\pi \qquad$ &  $\qquad \phi_{v_{d}}/\pi \qquad$ &  $\qquad \phi_{\upsilon_{\nu^{c}}}/\pi \qquad$ &  $\qquad M_{2} \qquad$ \\ \hline
  $[695;710]$ & [$-$0.024;$-$0.022] & [$-$0.4;$-$0.1] & [0.382;0.385] & [$-$0.095;$-$0.090] & [$-$1;$-$0.9] & 800 \\
  \hline
\end{tabular}
\caption{The parameter space of the random scan plot. All dimensionful parameters are given in GeV.}
\label{para1}
\end{table}

\begin{figure}
\setlength{\unitlength}{1mm}
\begin{minipage}[c]{0.5\textwidth}
\centering
\includegraphics[width=3in]{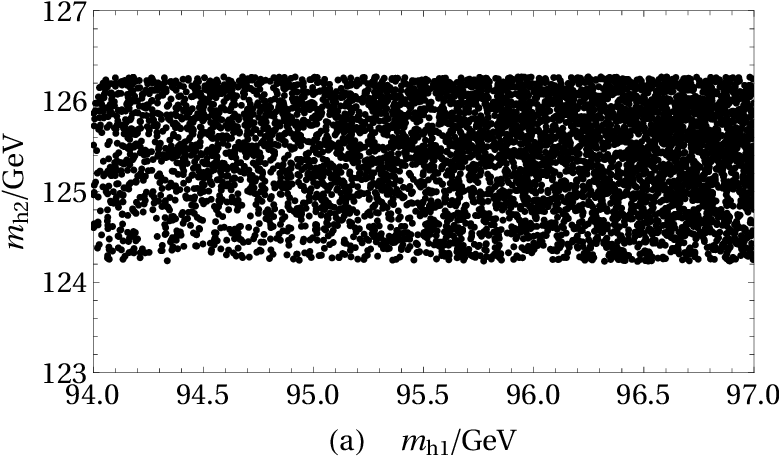}
\end{minipage}%
\begin{minipage}[c]{0.5\textwidth}
\centering
\includegraphics[width=3in]{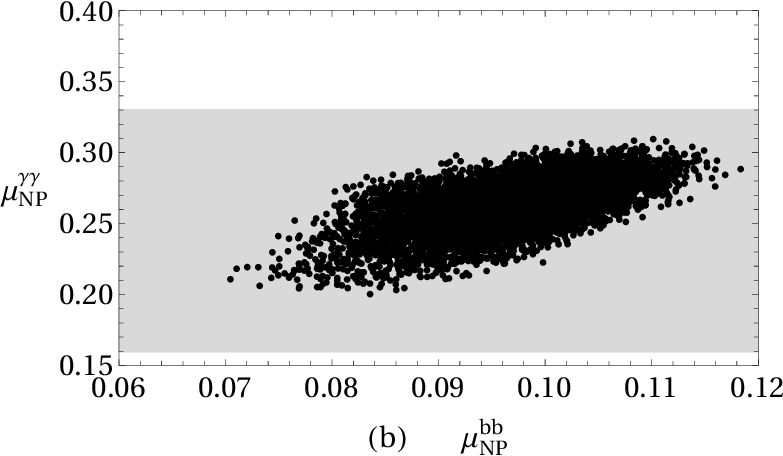}
\end{minipage}%
\caption[]{The left plot (a) shows the mass range of $h_1$ and $h_2$; for $h_2$, we take 1 GeV theory uncertainty, for $h_1$, CMS gives a mass of around 95.4 GeV, and we take 3 GeV experimental error. Correlation of these two signal strengths in (b); the gray area is the 1$\sigma$ experimental error. The values of parameters are in Table \ref{para1}.}
\label{gaga-bb}
\end{figure}
Let us remark that, in the random scan plots, we must first ensure that the mass of $h_2$ can conform to the experimental constraints, because the mass determination for SM-like Higgs is already very accurate. In Fig.\ref{gaga-bb} (a), we chose 1 GeV theory uncertainty for the SM-like Higgs. For $h_1$, we can choose the theory uncertainty up to 3 GeV, because CMS shows only that there is a 2.8 $\sigma$ excess at 95.3 GeV \cite{CMS1,CMS5,CMS9}, which is not exactly observed. In Fig \ref{gaga-bb}. (b), the $(\mu_{NP}^{\gamma\gamma},\mu_{NP}^{bb})$ plane, we can clearly see that with the parameter space of Table \ref{para1}, most points can explain the dphoton excess and the $b\bar{b}$ excess.

\section{Conclusion\label{sec5} }

In this paper we introduced CPV in the $\mu$-from-$\nu$SSM, which leads to $CP$-even Higgs sector mixed with $CP$-odd Higgs sector. We also analyzed an excess in the diphoton decay mode at $\sim$ 95 GeV as reported by ATLAS and CMS, together with a $\sim$ 2 $\sigma$ excess at LEP in the same mass range. The mixing and CPV are used to produce the lightest Higgs boson mass around 95 GeV, and the next-to-lightest Higgs boson mass around 125 GeV is the so-called SM-like Higgs. The lightest Higgs boson can explain an excess of $\gamma\gamma$ events at $\sim$ 96 GeV as reported by CMS.

In the numerical part, we find a suitable parameter space, based on which we show the properties of the lightest and next-to-lightest Higgs boson. We found it very easy get both $\mu_{NP}^{bb}$ and $\mu_{NP}^{\gamma\gamma}$ to reach the central values of their respective experiments at the same time. We also analyze the influence of relevant parameters and CPV phases on the signal strengths or Higgs masses, and improve the signal strengths as much as possible while ensuring that the SM-like Higgs meets the experimental constraints and the ``diphoton excess'' is around 95 GeV.
\begin{acknowledgments}
\indent\indent
The work has been supported by the National Natural Science Foundation of China (NNSFC) with Grants No. 12075074, No. 12235008, No. 11705045, and No. 11535002, Hebei Natural Science Foundation with Grants No. A2022201017 and No. A2023201041, Natural Science Foundation of Guangxi Autonomous Region with Grant No. 2022GXNSFDA035068, and the youth top-notch talent support program of the Hebei Province.
\end{acknowledgments}

\appendix

\section{mass matrices}

The $CP$-even neutral scalars have the composition $S^{T}=(h_d,h_u, (\tilde{\nu}_{i}^{c})^{\text{Re}})$, and one can write the mass matrix $M_{S}^{2}$ as
\begin{eqnarray}
M_{S}^{2}=
\left(
\begin{array}{ccccc}
M_{h_{d}h_{d}}^{2} & M_{h_{d}h_{u}}^{2} & M_{h_{d}(\tilde{\nu}_{1}^{c})^{\text{Re}}}^{2} & M_{h_{d}(\tilde{\nu}_{2}^{c})^{\text{Re}}}^{2} & M_{h_{d}(\tilde{\nu}_{3}^{c})^{\text{Re}}}^{2} \\
M_{h_{u}h_{d}}^{2} & M_{h_{u}h_{u}}^{2} & M_{h_{u}(\tilde{\nu}_{1}^{c})^{\text{Re}}}^{2} & M_{h_{u}(\tilde{\nu}_{2}^{c})^{\text{Re}}}^{2} & M_{h_{u}(\tilde{\nu}_{3}^{c})^{\text{Re}}}^{2} \\
M_{(\tilde{\nu}_{1}^{c})^{\text{Re}}h_{d}}^{2} & M_{(\tilde{\nu}_{1}^{c})^{\text{Re}}h_{u}}^{2} & M_{(\tilde{\nu}_{1}^{c})^{\text{Re}}(\tilde{\nu}_{1}^{c})^{\text{Re}}}^{2} & M_{(\tilde{\nu}_{1}^{c})^{\text{Re}}(\tilde{\nu}_{2}^{c})^{\text{Re}}}^{2} & M_{(\tilde{\nu}_{1}^{c})^{\text{Re}}(\tilde{\nu}_{3}^{c})^{\text{Re}}}^{2} \\
M_{(\tilde{\nu}_{2}^{c})^{\text{Re}}h_{d}}^{2} & M_{(\tilde{\nu}_{2}^{c})^{\text{Re}}h_{u}}^{2} & M_{(\tilde{\nu}_{2}^{c})^{\text{Re}}(\tilde{\nu}_{1}^{c})^{\text{Re}}}^{2} & M_{(\tilde{\nu}_{2}^{c})^{\text{Re}}(\tilde{\nu}_{2}^{c})^{\text{Re}}}^{2} & M_{(\tilde{\nu}_{2}^{c})^{\text{Re}}(\tilde{\nu}_{3}^{c})^{\text{Re}}}^{2} \\
M_{(\tilde{\nu}_{3}^{c})^{\text{Re}}h_{d}}^{2} & M_{(\tilde{\nu}_{3}^{c})^{\text{Re}}h_{u}}^{2} & M_{(\tilde{\nu}_{3}^{c})^{\text{Re}}(\tilde{\nu}_{1}^{c})^{\text{Re}}}^{2} & M_{(\tilde{\nu}_{3}^{c})^{\text{Re}}(\tilde{\nu}_{2}^{c})^{\text{Re}}}^{2} & M_{(\tilde{\nu}_{3}^{c})^{\text{Re}}(\tilde{\nu}_{3}^{c})^{\text{Re}}}^{2} \\
\end{array}
\right)
\end{eqnarray}
\begin{eqnarray}
M_{h_{d}h_{d}}^{2}\!\!\!\!\!&&={{G^2}\over{2}}v_{d}^2+\text{Re}(e^{i\phi_{\lambda}} e^{-i\phi_{v_{u}}} e^{i\phi_{v_d}} e^{i\phi_{\upsilon_{\nu_{i}^c}}})(A_{\lambda}\lambda)_{i}\upsilon_{\nu_{i}^{c}}\tan\beta \nonumber \\
&&+\text{Re}(e^{-i\phi_{\lambda}} e^{-i\phi_{v_{u}}} e^{i\phi_{v_d}} e^{i\phi_{\upsilon_{\nu_{i}^c}}} e^{i\phi_{\upsilon_{\nu_{j}^c}}}\lambda_{k}^{\ast}\kappa_{ijk})\upsilon_{\nu_{i}^{c}} \upsilon_{\nu_{j}^{c}} \tan\beta +\Delta_{11}, \\
M_{h_{d}h_{u}}^{2}\!\!\!\!\!&&=-{{G^2}\over{2}}v_{d}v_{u}-\text{Re}(e^{i\phi_{\lambda}} e^{-i\phi_{v_{u}}} e^{i\phi_{v_d}} e^{i\phi_{\upsilon_{\nu_{i}^c}}})(A_{\lambda}\lambda)_{i}\upsilon_{\nu_{i}^{c}}+2|\lambda_{i}|^2v_{d}v_{u} \nonumber \\
&&-\text{Re}(e^{-i\phi_{\lambda}} e^{-i\phi_{v_{u}}} e^{i\phi_{v_d}} e^{i\phi_{\upsilon_{\nu_{i}^c}}} e^{i\phi_{\upsilon_{\nu_{j}^c}}}\lambda_{k}^{\ast}\kappa_{ijk})\upsilon_{\nu_{i}^{c}} \upsilon_{\nu_{j}^{c}}+\Delta_{12}, \\
M_{h_{u}h_{u}}^{2}\!\!\!\!\!&&={{G^2}\over{2}}v_{u}^2+\text{Re}(e^{i\phi_{\lambda}} e^{-i\phi_{v_{u}}} e^{i\phi_{v_d}} e^{i\phi_{\upsilon_{\nu_{i}^c}}})(A_{\lambda}\lambda)_{i}\upsilon_{\nu_{i}^{c}}\cot\beta \nonumber \\
&&+\text{Re}(e^{-i\phi_{\lambda}} e^{-i\phi_{v_{u}}} e^{i\phi_{v_d}} e^{i\phi_{\upsilon_{\nu_{i}^c}}} e^{i\phi_{\upsilon_{\nu_{j}^c}}}\lambda_{k}^{\ast}\kappa_{ijk})\upsilon_{\nu_{i}^{c}} \upsilon_{\nu_{j}^{c}}\cot\beta+\Delta_{22}, \\
M_{h_{d}(\tilde{\nu}_{i}^{c})^{\text{Re}}}^{2}\!\!\!\!\!&&=-\text{Re}(e^{i\phi_{\lambda}} e^{-i\phi_{v_{u}}} e^{i\phi_{v_d}} e^{i\phi_{\upsilon_{\nu_{i}^c}}})(A_{\lambda}\lambda)_{i}v_{u}+2\lambda_{i}\lambda_{j}^{*} v_d \upsilon_{\nu_{j}^{c}} \nonumber \\
&&-2\text{Re}(e^{-i\phi_{\lambda}} e^{-i\phi_{v_{u}}} e^{i\phi_{v_d}} e^{i\phi_{\upsilon_{\nu_{i}^c}}} e^{i\phi_{\upsilon_{\nu_{j}^c}}}\lambda_{k}^{\ast}\kappa_{ijk})v_{u}\upsilon_{\nu_{j}^{c}}+\Delta_{1(2+i)}, \\
M_{h_{u}(\tilde{\nu}_{i}^{c})^{\text{Re}}}^{2}\!\!\!\!\!&&=-\text{Re}(e^{i\phi_{\lambda}} e^{-i\phi_{v_{u}}} e^{i\phi_{v_d}} e^{i\phi_{\upsilon_{\nu_{i}^c}}})(A_{\lambda}\lambda)_{i}v_{d}+2\lambda_{i}\lambda_{j}^{*} v_u \upsilon_{\nu_{j}^{c}} \nonumber \\
&&-2\text{Re}(e^{-i\phi_{\lambda}} e^{-i\phi_{v_{u}}} e^{i\phi_{v_d}} e^{i\phi_{\upsilon_{\nu_{i}^c}}} e^{i\phi_{\upsilon_{\nu_{j}^c}}}\lambda_{k}^{\ast}\kappa_{ijk})v_{d}\upsilon_{\nu_{j}^{c}}+\Delta_{2(2+i)}, \\
M_{(\tilde{\nu}_{i}^{c})^{\text{Re}}(\tilde{\nu}_{i}^{c})^{\text{Re}}}^{2}\!\!\!\!\!&&=(A_{\kappa} \kappa)_{ijk}\text{Re}(e^{i\phi_{\upsilon_{\nu_{k}^c}}} e^{i\phi_{\upsilon_{\nu_{k}^c}}} e^{i\phi_{\upsilon_{\nu_{k}^c}}})\upsilon_{\nu_{k}^c}+\Delta_{(2+i)(2+j)} \nonumber \\
&&+\text{Re}(e^{i\phi_{\lambda}} e^{-i\phi_{v_{u}}} e^{i\phi_{v_d}} e^{i\phi_{\upsilon_{\nu_{i}^c}}})(A_{\lambda}\lambda)_{i}{{v_d v_u}\over{\upsilon_{\nu_{j}^c}}},
\\%
M_{(\tilde{\nu}_{i}^{c})^{\text{Re}}(\tilde{\nu}_{j}^{c})^{\text{Re}}}^{2}\!\!\!\!\!&&=|\lambda_{i}|^2(v_{d}^2+v_{u}^{2})+\Delta_{(2+i)(2+j)}.
\end{eqnarray}

The radiative corrections from the top quark and bottom quark and their corresponding supersymmetric partners, the corrections in the mass matrix, can be expressed as
\begin{eqnarray}
\Delta_{ab}=\Delta_{ab}^t+\Delta_{ab}^b
\end{eqnarray}
\begin{eqnarray}
\Delta_{1(2+i)}^{t}\!\!\!\!\!&&={{3G_{F}m_{t}^{4}}\over{2\sqrt{2}\pi^2\sin^2\beta}} \Bigg\{ {{  {1\over2} \upsilon_{\nu_{j}^c} v_{d}(\lambda_{i}\lambda_{j}^{*}+\lambda_{i}^{*}\lambda_{j})\cot\beta-\text{Re}(e^{i\phi_{\lambda}} e^{i\phi_{A_{t}}} e^{i\phi_{v_{u}}} e^{-i\phi_{v_d}} e^{i\phi_{\upsilon_{\nu_{i}^c}}}) A_{t}\lambda_{i}v_{d}  }\over{ m_{\tilde{t}_{1}}^2-m_{\tilde{t}_{2}}^2 }} \nonumber \\
&&\times  {{|\mu|^2\cot\beta-\text{Re}(e^{i\phi_{\lambda}} e^{i\phi_{A_{t}}} e^{i\phi_{v_{u}}} e^{-i\phi_{v_d}} e^{i\phi_{\upsilon_{\nu_{i}^c}}})A_{t}\mu  }\over{ m_{\tilde{t}_{1}}^2-m_{\tilde{t}_{2}}^2 }} g(m_{\tilde{t}_{1}}^2,m_{\tilde{t}_{2}}^2)  \Bigg\},   \\
\Delta_{2(2+i)}^{t}\!\!\!\!\!&&={{3G_{F}m_{t}^{4}}\over{2\sqrt{2}\pi^2\sin^2\beta}} \Bigg\{ {{  {1\over2} \upsilon_{\nu_{j}^c} v_{d}(\lambda_{i}\lambda_{j}^{*}+\lambda_{i}^{*}\lambda_{j})\cot\beta-\text{Re}(e^{i\phi_{\lambda}} e^{i\phi_{A_{t}}} e^{i\phi_{v_{u}}} e^{-i\phi_{v_d}} e^{i\phi_{\upsilon_{\nu_{i}^c}}}) A_{t}\lambda_{i}v_{d}  }\over{ m_{\tilde{t}_{1}}^2-m_{\tilde{t}_{2}}^2 }} \Bigg\} \nonumber \\
&&\times \Bigg\{ \ln{{m_{\tilde{t}_{1}}^2}\over{m_{\tilde{t}_{2}}^2}} + {{|A_{t}|^2-\text{Re}(e^{i\phi_{\lambda}} e^{i\phi_{A_{t}}} e^{i\phi_{v_{u}}} e^{-i\phi_{v_d}} e^{i\phi_{\upsilon_{\nu_{i}^c}}})A_{t}\mu\cot\beta  }\over{ m_{\tilde{t}_{1}}^2-m_{\tilde{t}_{2}}^2 }} g(m_{\tilde{t}_{1}}^2,m_{\tilde{t}_{2}}^2)    \Bigg\}, \\
\Delta_{(2+i)(2+j)}^{t}\!\!\!\!\!&&={{3G_{F}m_{t}^{4}}\over{2\sqrt{2}\pi^2\sin^2\beta}} \Bigg\{ {{  ({1\over2} \upsilon_{\nu_{j}^c} v_{d}(\lambda_{i}\lambda_{j}^{*}+\lambda_{i}^{*}\lambda_{j})\cot\beta-\text{Re}(e^{i\phi_{\lambda}} e^{i\phi_{A_{t}}} e^{i\phi_{v_{u}}} e^{-i\phi_{v_d}} e^{i\phi_{\upsilon_{\nu_{i}^c}}}) A_{t}\lambda_{i}v_{d})^2  }\over{ (m_{\tilde{t}_{1}}^2-m_{\tilde{t}_{2}}^2)^2 }} \nonumber \\
&&\times g(m_{\tilde{t}_{1}}^2,m_{\tilde{t}_{2}}^2) \Bigg\}
\end{eqnarray}
\begin{eqnarray}
\Delta_{11}^{b}\!\!\!\!\!&&={{3G_{F}m_{b}^{4}}\over{2\sqrt{2}\pi^2\cos^2\beta}} \Bigg\{\ln{{m_{\tilde{b}_{1}}^2 m_{\tilde{b}_{2}}^2 }\over{m_{b}^{4}}} +{{2|A_{b}|^2-\text{Re}(e^{i\phi_{\lambda}} e^{i\phi_{v_{u}}} e^{-i\phi_{v_d}} e^{i\phi_{\upsilon_{\nu_{i}^c}}}) A_{b}\mu\tan\beta }\over{m_{\tilde{b}_{1}}^2-m_{\tilde{b}_{2}}^2}} \ln{{m_{\tilde{b}_{1}}^2}\over{m_{\tilde{b}_{2}}^2}} \nonumber \\
&&+{{( |A_{b}|^2-\text{Re}(e^{i\phi_{\lambda}} e^{i\phi_{v_{u}}} e^{-i\phi_{v_d}} e^{i\phi_{\upsilon_{\nu_{i}^c}}}) A_{b}\mu\tan\beta )^2   }\over{ (m_{\tilde{b}_{1}}^2-m_{\tilde{b}_{2}}^2)^2 }} g(m_{\tilde{b}_{1}}^2,m_{\tilde{b}_{2}}^2)  \Bigg\}, \\
\Delta_{12}^{b}\!\!\!\!\!&&={{3G_{F}m_{b}^{4}}\over{2\sqrt{2}\pi^2\cos^2\beta}} \Bigg\{ {{|\mu|^2\tan\beta-\text{Re}(e^{i\phi_{\lambda}} e^{i\phi_{v_{u}}} e^{-i\phi_{v_d}} e^{i\phi_{\upsilon_{\nu_{i}^c}}}) A_{b}\mu  }\over{ m_{\tilde{b}_{1}}^2-m_{\tilde{b}_{2}}^2 }} \Bigg\} \nonumber \\
&&\times \Bigg\{\ln{{m_{\tilde{b}_{1}}^2}\over{m_{\tilde{b}_{2}}^2}}+{{|A_{b}|^2-\text{Re}(e^{i\phi_{\lambda}} e^{i\phi_{v_{u}}} e^{-i\phi_{v_d}} e^{i\phi_{\upsilon_{\nu_{i}^c}}}) A_{b}\mu\tan\beta }\over{ m_{\tilde{b}_{1}}^2-m_{\tilde{b}_{2}}^2 }} g(m_{\tilde{b}_{1}}^2,m_{\tilde{b}_{2}}^2)    \Bigg\}, \\
\Delta_{22}^{b}\!\!\!\!\!&&={{3G_{F}m_{b}^{4}}\over{2\sqrt{2}\pi^2\cos^2\beta}} \Bigg\{ {{(|\mu|^2\tan\beta-\text{Re}(e^{i\phi_{\lambda}} e^{i\phi_{v_{u}}} e^{-i\phi_{v_d}} e^{i\phi_{\upsilon_{\nu_{i}^c}}}) A_{b}\mu)^2  }\over{ (m_{\tilde{b}_{1}}^2-m_{\tilde{b}_{2}}^2)^2 }} g(m_{\tilde{b}_{1}}^2,m_{\tilde{b}_{2}}^2),  \Bigg\} \\
\Delta_{1(2+i)}^{b}\!\!\!\!\!&&={{3G_{F}m_{b}^{4}}\over{2\sqrt{2}\pi^2\cos^2\beta}} \Bigg\{ {{ {1\over2} \upsilon_{\nu_{j}^c} v_{u}(\lambda_{i}\lambda_{j}^{*}+\lambda_{i}^{*}\lambda_{j})\tan\beta-\text{Re}(e^{i\phi_{\lambda}} e^{i\phi_{v_{u}}} e^{-i\phi_{v_d}} e^{i\phi_{\upsilon_{\nu_{i}^c}}}) A_{b}\lambda_{i}v_{u} }\over{ m_{\tilde{b}_{1}}^2-m_{\tilde{b}_{2}}^2 }} \Bigg\} \nonumber \\
&&\times \Bigg\{\ln{{m_{\tilde{b}_{1}}^2}\over{m_{\tilde{b}_{2}}^2}}+{{|A_{b}|^2-\text{Re}(e^{i\phi_{\lambda}} e^{i\phi_{v_{u}}} e^{-i\phi_{v_d}} e^{i\phi_{\upsilon_{\nu_{i}^c}}}) A_{b}\mu\tan\beta }\over{ m_{\tilde{b}_{1}}^2-m_{\tilde{b}_{2}}^2 }} g(m_{\tilde{b}_{1}}^2,m_{\tilde{b}_{2}}^2) \Bigg\}, \\
\Delta_{2(2+i)}^{b}\!\!\!\!\!&&={{3G_{F}m_{b}^{4}}\over{2\sqrt{2}\pi^2\cos^2\beta}} \Bigg\{{{|\mu|^2\tan\beta-\text{Re}(e^{i\phi_{\lambda}} e^{i\phi_{v_{u}}} e^{-i\phi_{v_d}} e^{i\phi_{\upsilon_{\nu_{i}^c}}}) A_{b}\mu  }\over{ m_{\tilde{b}_{1}}^2-m_{\tilde{b}_{2}}^2 }} g(m_{\tilde{b}_{1}}^2,m_{\tilde{b}_{2}}^2) \nonumber \\
&&\times {{  {1\over2} \upsilon_{\nu_{j}^c} v_{u}(\lambda_{i}\lambda_{j}^{*}+\lambda_{i}^{*}\lambda_{j})\tan\beta-\text{Re}(e^{i\phi_{\lambda}} e^{i\phi_{v_{u}}} e^{-i\phi_{v_d}} e^{i\phi_{\upsilon_{\nu_{i}^c}}}) A_{b}\lambda_{i}v_{u}  }\over{ m_{\tilde{b}_{1}}^2-m_{\tilde{b}_{2}}^2 }} \Bigg\} \\
\Delta_{(2+i)(2+i)}^{b}\!\!\!\!\!&&={{3G_{F}m_{b}^{4}}\over{2\sqrt{2}\pi^2\cos^2\beta}} {{  ({1\over2} \upsilon_{\nu_{j}^c} v_{u}(\lambda_{i}\lambda_{j}^{*}+\lambda_{i}^{*}\lambda_{j})\tan\beta-\text{Re}(e^{i\phi_{\lambda}} e^{i\phi_{v_{u}}} e^{-i\phi_{v_d}} e^{i\phi_{\upsilon_{\nu_{i}^c}}}) A_{b}\lambda_{i}v_{u})^2  }\over{ (m_{\tilde{b}_{1}}^2-m_{\tilde{b}_{2}}^2)^2 }} \nonumber \\
&&\times g(m_{\tilde{b}_{1}}^2,m_{\tilde{b}_{2}}^2).
\end{eqnarray}

In the same way, the $CP$-odd neutral scalars mass matrix is that
\begin{eqnarray}
M_{P}^{2}=
\left(
\begin{array}{ccccc}
M_{\sigma_{d}\sigma_{d}}^{2} & M_{\sigma_{d}\sigma_{u}}^{2} & M_{\sigma_{d}(\tilde{\nu}_{1}^{c})^{\text{Im}}}^{2} & M_{\sigma_{d}(\tilde{\nu}_{2}^{c})^{\text{Im}}}^{2} & M_{\sigma_{d}(\tilde{\nu}_{3}^{c})^{\text{Im}}}^{2} \\
M_{\sigma_{u}\sigma_{d}}^{2} & M_{\sigma_{u}\sigma_{u}}^{2} & M_{\sigma_{u}(\tilde{\nu}_{1}^{c})^{\text{Im}}}^{2} & M_{\sigma_{u}(\tilde{\nu}_{2}^{c})^{\text{Im}}}^{2} & M_{\sigma_{u}(\tilde{\nu}_{3}^{c})^{\text{Im}}}^{2} \\
M_{(\tilde{\nu}_{1}^{c})^{\text{Im}}\sigma_{d}}^{2} & M_{(\tilde{\nu}_{1}^{c})^{\text{Im}}\sigma_{u}}^{2} & M_{(\tilde{\nu}_{1}^{c})^{\text{Im}}(\tilde{\nu}_{1}^{c})^{\text{Im}}}^{2} & M_{(\tilde{\nu}_{1}^{c})^{\text{Im}}(\tilde{\nu}_{2}^{c})^{\text{Im}}}^{2} & M_{(\tilde{\nu}_{1}^{c})^{\text{Im}}(\tilde{\nu}_{3}^{c})^{\text{Im}}}^{2} \\
M_{(\tilde{\nu}_{2}^{c})^{\text{Im}}\sigma_{d}}^{2} & M_{(\tilde{\nu}_{2}^{c})^{\text{Im}}\sigma_{u}}^{2} & M_{(\tilde{\nu}_{2}^{c})^{\text{Im}}(\tilde{\nu}_{1}^{c})^{\text{Im}}}^{2} & M_{(\tilde{\nu}_{2}^{c})^{\text{Im}}(\tilde{\nu}_{2}^{c})^{\text{Im}}}^{2} & M_{(\tilde{\nu}_{2}^{c})^{\text{Im}}(\tilde{\nu}_{3}^{c})^{\text{Im}}}^{2} \\
M_{(\tilde{\nu}_{3}^{c})^{\text{Im}}\sigma_{d}}^{2} & M_{(\tilde{\nu}_{3}^{c})^{\text{Im}}\sigma_{u}}^{2} & M_{(\tilde{\nu}_{3}^{c})^{\text{Im}}(\tilde{\nu}_{1}^{c})^{\text{Im}}}^{2} & M_{(\tilde{\nu}_{3}^{c})^{\text{Im}}(\tilde{\nu}_{2}^{c})^{\text{Im}}}^{2} & M_{(\tilde{\nu}_{3}^{c})^{\text{Im}}(\tilde{\nu}_{3}^{c})^{\text{Im}}}^{2} \\
\end{array}
\right)
\end{eqnarray}
\begin{eqnarray}
M_{\sigma_{d}\sigma_{d}}^{2}\!\!\!\!\!&&=\text{Re}(e^{i\phi_{\lambda}} e^{-i\phi_{v_{u}}} e^{i\phi_{v_d}} e^{i\phi_{\upsilon_{\nu_{i}^c}}})(A_{\lambda}\lambda)_{i}\upsilon_{\nu_{i}^c}\tan\beta \nonumber \\
&&+\text{Re}(e^{-i\phi_{\lambda}} e^{-i\phi_{v_{u}}} e^{i\phi_{v_d}} e^{i\phi_{\upsilon_{\nu_{i}^c}}} e^{i\phi_{\upsilon_{\nu_{j}^c}}}\lambda_{k}^{\ast}\kappa_{ijk}) \upsilon_{\nu_{i}^c} \upsilon_{\nu_{j}^c} \tan\beta+\Delta_{66}, \\
M_{\sigma_{d}\sigma_{u}}^{2}\!\!\!\!\!&&=-\text{Re}(e^{i\phi_{\lambda}} e^{-i\phi_{v_{u}}} e^{i\phi_{v_d}} e^{i\phi_{\upsilon_{\nu_{i}^c}}})(A_{\lambda}\lambda)_{i} \upsilon_{\nu_{i}^c} \nonumber \\
&&+\text{Re}(e^{-i\phi_{\lambda}} e^{-i\phi_{v_{u}}} e^{i\phi_{v_d}} e^{i\phi_{\upsilon_{\nu_{i}^c}}} e^{i\phi_{\upsilon_{\nu_{j}^c}}}\lambda_{k}^{\ast}\kappa_{ijk}) \upsilon_{\nu_{i}^c} \upsilon_{\nu_{j}^c} +\Delta_{67}, \\
M_{\sigma_{u}\sigma_{u}}^{2}\!\!\!\!\!&&=\text{Re}(e^{i\phi_{\lambda}} e^{-i\phi_{v_{u}}} e^{i\phi_{v_d}} e^{i\phi_{\upsilon_{\nu_{i}^c}}})(A_{\lambda}\lambda)_{i} \upsilon_{\nu_{i}^c} \cot\beta \nonumber \\
&&+\text{Re}(e^{-i\phi_{\lambda}} e^{-i\phi_{v_{u}}} e^{i\phi_{v_d}} e^{i\phi_{\upsilon_{\nu_{i}^c}}} e^{i\phi_{\upsilon_{\nu_{j}^c}}}\lambda_{k}^{\ast}\kappa_{ijk}) \upsilon_{\nu_{i}^c} \upsilon_{\nu_{j}^c} \cot\beta+\Delta_{77}, \\
M_{\sigma_{d}(\tilde{\nu}_{i}^{c})^{\text{Im}}}^{2}\!\!\!\!\!&&=\text{Re}(e^{i\phi_{\lambda}} e^{-i\phi_{v_{u}}} e^{i\phi_{v_d}} e^{i\phi_{\upsilon_{\nu_{i}^c}}})(A_{\lambda}\lambda)_{i}v_{u} \nonumber \\
&&+2\text{Re}(e^{-i\phi_{\lambda}} e^{-i\phi_{v_{u}}} e^{i\phi_{v_d}} e^{i\phi_{\upsilon_{\nu_{i}^c}}} e^{i\phi_{\upsilon_{\nu_{j}^c}}}\lambda_{k}^{\ast}\kappa_{ijk}) v_{u} \upsilon_{\nu_{i}^c}+\Delta_{6(7+i)}, \\
M_{\sigma_{u}(\tilde{\nu}_{i}^{c})^{\text{Im}}}^{2}\!\!\!\!\!&&=-\text{Re}(e^{i\phi_{\lambda}} e^{-i\phi_{v_{u}}} e^{i\phi_{v_d}} e^{i\phi_{\upsilon_{\nu_{i}^c}}})(A_{\lambda}\lambda)_{i}v_{d} \nonumber \\
&&-2\text{Re}(e^{-i\phi_{\lambda}} e^{-i\phi_{v_{u}}} e^{i\phi_{v_d}} e^{i\phi_{\upsilon_{\nu_{i}^c}}} e^{i\phi_{\upsilon_{\nu_{j}^c}}}\lambda_{k}^{\ast}\kappa_{ijk}) v_{d} \upsilon_{\nu_{i}^c}+\Delta_{7(7+i)}, \\
M_{(\tilde{\nu}_{i}^{c})^{\text{Im}}(\tilde{\nu}_{i}^{c})^{\text{Im}}}^{2}\!\!\!\!\!&&=\text{Re}(e^{i\phi_{\lambda}} e^{-i\phi_{v_{u}}} e^{i\phi_{v_d}} e^{i\phi_{\upsilon_{\nu_{i}^c}}}) (A_{\lambda}\lambda)_{i} {{v_{d}v_{u}}\over{\upsilon_{\nu_{i}^c}}}-(A_{\kappa}\kappa)_{ijk} \text{Re}(e^{i\phi_{\upsilon_{\nu_{i}^c}}} e^{i\phi_{\upsilon_{\nu_{i}^c}}} e^{i\phi_{\upsilon_{\nu_{i}^c}}}) \upsilon_{\nu_{k}^c} \nonumber \\
&&+4\text{Re}(e^{-i\phi_{\lambda}} e^{-i\phi_{v_{u}}} e^{i\phi_{v_d}} e^{i\phi_{\upsilon_{\nu_{i}^c}}} e^{i\phi_{\upsilon_{\nu_{j}^c}}}\lambda_{k}^{\ast}\kappa_{ijk}) v_{d} v_{u}+\Delta_{(7+i)(7+i)}, \\
M_{(\tilde{\nu}_{i}^{c})^{\text{Im}}(\tilde{\nu}_{j}^{c})^{\text{Im}}}^{2}\!\!\!\!\!&&=|\lambda_{i}|^2(v_{u}^2+v_{d}^2)+\Delta_{(7+i)(7+i)},
\end{eqnarray}

\begin{eqnarray}
\Delta_{66}^{t}\!\!\!\!\!&&={{3G_{F}m_{t}^{4}}\over{2\sqrt{2}\pi^2\sin^2\beta}} {{ (-\text{Im}(e^{i\phi_{\lambda}} e^{i\phi_{A_{t}}} e^{i\phi_{v_{u}}} e^{-i\phi_{v_d}} e^{i\phi_{\upsilon_{\nu_{i}^c}}})A_{t}\mu)^2   }\over{(m_{\tilde{t}_{1}}^2-m_{\tilde{t}_{2}}^2})^2} g(m_{\tilde{t}_{1}}^2,m_{\tilde{t}_{2}}^2) , \\
\Delta_{67}^{t}\!\!\!\!\!&&={{3G_{F}m_{t}^{4}}\over{2\sqrt{2}\pi^2\sin^2\beta}} {{\text{Im}(e^{i\phi_{\lambda}} e^{i\phi_{A_{t}}} e^{i\phi_{v_{u}}} e^{-i\phi_{v_d}} e^{i\phi_{\upsilon_{\nu_{i}^c}}})A_{t}\mu \cot\beta  }\over{m_{\tilde{t}_{1}}^2-m_{\tilde{t}_{2}}^2}} \nonumber \\
&&\times {{ -\text{Im}(e^{i\phi_{\lambda}} e^{i\phi_{A_{t}}} e^{i\phi_{v_{u}}} e^{-i\phi_{v_d}} e^{i\phi_{\upsilon_{\nu_{i}^c}}})A_{t}\mu }\over{m_{\tilde{t}_{1}}^2-m_{\tilde{t}_{2}}^2}} g(m_{\tilde{t}_{1}}^2,m_{\tilde{t}_{2}}^2),  \\
\Delta_{77}^{t}\!\!\!\!\!&&={{3G_{F}m_{t}^{4}}\over{2\sqrt{2}\pi^2\sin^2\beta}} { {(\text{Im}(e^{i\phi_{\lambda}} e^{i\phi_{A_{t}}} e^{i\phi_{v_{u}}} e^{-i\phi_{v_d}} e^{i\phi_{\upsilon_{\nu_{i}^c}}})A_{t}\mu \cot\beta)^2  }\over{(m_{\tilde{t}_{1}}^2-m_{\tilde{t}_{2}}^2)^2} } g(m_{\tilde{t}_{1}}^2,m_{\tilde{t}_{2}}^2), \\
\Delta_{6(7+i)}^{t}\!\!\!\!\!&&={{3G_{F}m_{t}^{4}}\over{2\sqrt{2}\pi^2\sin^2\beta}} {{ -\text{Im}(e^{i\phi_{\lambda}} e^{i\phi_{A_{t}}} e^{i\phi_{v_{u}}} e^{-i\phi_{v_d}} e^{i\phi_{\upsilon_{\nu_{i}^c}}})A_{t}\mu }\over{m_{\tilde{t}_{1}}^2-m_{\tilde{t}_{2}}^2}} \nonumber \\
&&\times {{ \text{Im}(e^{i\phi_{\lambda}} e^{i\phi_{A_{t}}} e^{i\phi_{v_{u}}} e^{-i\phi_{v_d}} e^{i\phi_{\upsilon_{\nu_{i}^c}}})A_{t}\lambda_{i}v_{d} }\over{m_{\tilde{t}_{1}}^2-m_{\tilde{t}_{2}}^2}} g(m_{\tilde{t}_{1}}^2,m_{\tilde{t}_{2}}^2), \\
\Delta_{7(7+i)}^{t}\!\!\!\!\!&&={{3G_{F}m_{t}^{4}}\over{2\sqrt{2}\pi^2\sin^2\beta}} { {\text{Im}(e^{i\phi_{\lambda}} e^{i\phi_{A_{t}}} e^{i\phi_{v_{u}}} e^{-i\phi_{v_d}} e^{i\phi_{\upsilon_{\nu_{i}^c}}})A_{t}\mu \cot\beta  }\over{m_{\tilde{t}_{1}}^2-m_{\tilde{t}_{2}}^2} } \nonumber \\
&&\times { {-\text{Im}(e^{i\phi_{\lambda}} e^{i\phi_{A_{t}}} e^{i\phi_{v_{u}}} e^{-i\phi_{v_d}} e^{i\phi_{\upsilon_{\nu_{i}^c}}}) A_{t}\lambda_{i}v_{d}  }\over{m_{\tilde{t}_{1}}^2-m_{\tilde{t}_{2}}^2} } g(m_{\tilde{t}_{1}}^2,m_{\tilde{t}_{2}}^2), \\
\Delta_{(7+i)(7+j)}^{t}\!\!\!\!\!&&={{3G_{F}m_{t}^{4}}\over{2\sqrt{2}\pi^2\sin^2\beta}} { {(-\text{Im}(e^{i\phi_{\lambda}} e^{i\phi_{A_{t}}} e^{i\phi_{v_{u}}} e^{-i\phi_{v_d}} e^{i\phi_{\upsilon_{\nu_{i}^c}}}) A_{t}\lambda_{i}v_{d})^2  }\over{ (m_{\tilde{t}_{1}}^2-m_{\tilde{t}_{2}}^2})^2 } g(m_{\tilde{t}_{1}}^2,m_{\tilde{t}_{2}}^2)
\end{eqnarray}
\begin{eqnarray}
\Delta_{66}^{b}\!\!\!\!\!&&={{3G_{F}m_{b}^{4}}\over{2\sqrt{2}\pi^2\cos^2\beta}} {{(-\text{Im}(e^{i\phi_{\lambda}} e^{i\phi_{v_{u}}} e^{-i\phi_{v_d}} e^{i\phi_{\upsilon_{\nu_{i}^c}}}) A_{b}\mu\tan\beta)^2}\over{(m_{\tilde{b}_{1}}^2-m_{\tilde{b}_{2}}^2})^2} g(m_{\tilde{b}_{1}}^2,m_{\tilde{b}_{2}}^2), \\
\Delta_{67}^{b}\!\!\!\!\!&&={{3G_{F}m_{b}^{4}}\over{2\sqrt{2}\pi^2\cos^2\beta}} { {-(\text{Im}(e^{i\phi_{\lambda}} e^{i\phi_{v_{u}}} e^{-i\phi_{v_d}} e^{i\phi_{\upsilon_{\nu_{i}^c}}}) A_{b}\mu)^2 \tan\beta   }\over{(m_{\tilde{b}_{1}}^2-m_{\tilde{b}_{2}}^2)^2} } g(m_{\tilde{b}_{1}}^2,m_{\tilde{b}_{2}}^2), \\
\Delta_{77}^{b}\!\!\!\!\!&&={{3G_{F}m_{b}^{4}}\over{2\sqrt{2}\pi^2\cos^2\beta}} { {(\text{Im}(e^{i\phi_{\lambda}} e^{i\phi_{v_{u}}} e^{-i\phi_{v_d}} e^{i\phi_{\upsilon_{\nu_{i}^c}}}) A_{b}\mu)^2   }\over{(m_{\tilde{b}_{1}}^2-m_{\tilde{b}_{2}}^2)^2} } g(m_{\tilde{b}_{1}}^2,m_{\tilde{b}_{2}}^2),\\
\Delta_{6(7+i)}^{b}\!\!\!\!\!&&={{3G_{F}m_{b}^{4}}\over{2\sqrt{2}\pi^2\cos^2\beta}} { {-(\text{Im}(e^{i\phi_{\lambda}} e^{i\phi_{v_{u}}} e^{-i\phi_{v_d}} e^{i\phi_{\upsilon_{\nu_{i}^c}}}) A_{b})^2 \mu\tan\beta\lambda_{i} v_{u}  }\over{(m_{\tilde{b}_{1}}^2-m_{\tilde{b}_{2}}^2)^2} } g(m_{\tilde{b}_{1}}^2,m_{\tilde{b}_{2}}^2),\\
\Delta_{7(7+i)}^{b}\!\!\!\!\!&&={{3G_{F}m_{b}^{4}}\over{2\sqrt{2}\pi^2\cos^2\beta}} { {(\text{Im}(e^{i\phi_{\lambda}} e^{i\phi_{v_{u}}} e^{-i\phi_{v_d}} e^{i\phi_{\upsilon_{\nu_{i}^c}}}) A_{b})^2 \mu\lambda_{i} v_{u}  }\over{(m_{\tilde{b}_{1}}^2-m_{\tilde{b}_{2}}^2)^2} } g(m_{\tilde{b}_{1}}^2,m_{\tilde{b}_{2}}^2),\\
\Delta_{(7+i)(7+j)}^{b}\!\!\!\!\!&&={{3G_{F}m_{b}^{4}}\over{2\sqrt{2}\pi^2\cos^2\beta}} { {(\text{Im}(e^{i\phi_{\lambda}} e^{i\phi_{v_{u}}} e^{-i\phi_{v_d}} e^{i\phi_{\upsilon_{\nu_{i}^c}}}) A_{b}\lambda_{i} v_{u} )^2  }\over{(m_{\tilde{b}_{1}}^2-m_{\tilde{b}_{2}}^2)^2} } g(m_{\tilde{b}_{1}}^2,m_{\tilde{b}_{2}}^2).
\end{eqnarray}

The mix mass submatrix $M_{SP}^{2}$
\begin{eqnarray}
M_{SP}^{2}=
\left(
\begin{array}{ccccc}
M_{h_{d}\sigma_{d}}^{2} & M_{h_{d}\sigma_{u}}^{2} & M_{h_{d}(\tilde{\nu}_{1}^{c})^{\text{Im}}}^{2} & M_{h_{d}(\tilde{\nu}_{2}^{c})^{\text{Im}}}^{2} & M_{h_{d}(\tilde{\nu}_{3}^{c})^{\text{Im}}}^{2} \\
M_{h_{u}\sigma_{d}}^{2} & M_{h_{u}\sigma_{u}}^{2} & M_{h_{u}(\tilde{\nu}_{1}^{c})^{\text{Im}}}^{2} & M_{h_{u}(\tilde{\nu}_{2}^{c})^{\text{Im}}}^{2} & M_{h_{u}(\tilde{\nu}_{3}^{c})^{\text{Im}}}^{2} \\
M_{(\tilde{\nu}_{1}^{c})^{\text{Re}}\sigma_{d}}^{2} & M_{(\tilde{\nu}_{1}^{c})^{\text{Re}}\sigma_{u}}^{2} & M_{(\tilde{\nu}_{1}^{c})^{\text{Re}}(\tilde{\nu}_{1}^{c})^{\text{Im}}}^{2} & M_{(\tilde{\nu}_{1}^{c})^{\text{Re}}(\tilde{\nu}_{2}^{c})^{\text{Im}}}^{2} & M_{(\tilde{\nu}_{1}^{c})^{\text{Re}}(\tilde{\nu}_{3}^{c})^{\text{Im}}}^{2} \\
M_{(\tilde{\nu}_{2}^{c})^{\text{Re}}\sigma_{d}}^{2} & M_{(\tilde{\nu}_{2}^{c})^{\text{Re}}\sigma_{u}}^{2} & M_{(\tilde{\nu}_{2}^{c})^{\text{Re}}(\tilde{\nu}_{1}^{c})^{\text{Im}}}^{2} & M_{(\tilde{\nu}_{2}^{c})^{\text{Re}}(\tilde{\nu}_{2}^{c})^{\text{Im}}}^{2} & M_{(\tilde{\nu}_{2}^{c})^{\text{Re}}(\tilde{\nu}_{3}^{c})^{\text{Im}}}^{2} \\
M_{(\tilde{\nu}_{3}^{c})^{\text{Re}}\sigma_{d}}^{2} & M_{(\tilde{\nu}_{3}^{c})^{\text{Re}}\sigma_{u}}^{2} & M_{(\tilde{\nu}_{3}^{c})^{\text{Re}}(\tilde{\nu}_{1}^{c})^{\text{Im}}}^{2} & M_{(\tilde{\nu}_{3}^{c})^{\text{Re}}(\tilde{\nu}_{2}^{c})^{\text{Im}}}^{2} & M_{(\tilde{\nu}_{3}^{c})^{\text{Re}}(\tilde{\nu}_{3}^{c})^{\text{Im}}}^{2} \\
\end{array}
\right)
\end{eqnarray}

\begin{eqnarray}
M_{h_{d}\sigma_{d}}^{2}\!\!\!\!\!&&=\Delta_{16},  \\
M_{h_{d}\sigma_{u}}^{2}\!\!\!\!\!&&=\Delta_{17},  \\
M_{h_{u}\sigma_{d}}^{2}\!\!\!\!\!&&=\Delta_{26}, \\
M_{h_{u}\sigma_{u}}^{2}\!\!\!\!\!&&=\Delta_{27}, \\
M_{h_{d}(\tilde{\nu}_{i}^{c})^{\text{Im}}}^{2}\!\!\!\!\!&&=-\text{Im}(e^{i\phi_{\lambda}} e^{-i\phi_{v_{u}}} e^{i\phi_{v_d}} e^{i\phi_{\upsilon_{\nu_{i}^c}}}) (A_{\lambda}\lambda)_{i} v_{u}+\Delta_{1(7+i)},  \\
M_{h_{u}(\tilde{\nu}_{i}^{c})^{\text{Im}}}^{2}\!\!\!\!\!&&=-\text{Im}(e^{i\phi_{\lambda}} e^{-i\phi_{v_{u}}} e^{i\phi_{v_d}} e^{i\phi_{\upsilon_{\nu_{i}^c}}}) (A_{\lambda}\lambda)_{i} v_{d}+\Delta_{2(7+i)},  \\
M_{(\tilde{\nu}_{i}^{c})^{\text{Re}}\sigma_{d}}^{2}\!\!\!\!\!&&=-\text{Im}(e^{i\phi_{\lambda}} e^{-i\phi_{v_{u}}} e^{i\phi_{v_d}} e^{i\phi_{\upsilon_{\nu_{i}^c}}}) (A_{\lambda}\lambda)_{i} v_{u}+\Delta_{(2+i)6}, \\
M_{(\tilde{\nu}_{i}^{c})^{\text{Re}}\sigma_{u}}^{2}\!\!\!\!\!&&=\text{Im}(e^{i\phi_{\lambda}} e^{-i\phi_{v_{u}}} e^{i\phi_{v_d}} e^{i\phi_{\upsilon_{\nu_{i}^c}}}) (A_{\lambda}\lambda)_{i} v_{d}+\Delta_{(2+i)7}, \\
M_{(\tilde{\nu}_{i}^{c})^{\text{Re}}(\tilde{\nu}_{i}^{c})^{\text{Im}}}^{2}\!\!\!\!\!&&=2\text{Im}(e^{-i\phi_{\lambda}} e^{-i\phi_{v_{u}}} e^{i\phi_{v_d}} e^{i\phi_{\upsilon_{\nu_{i}^c}}} e^{i\phi_{\upsilon_{\nu_{j}^c}}}\lambda_{k}^{\ast}\kappa_{ijk}) v_{d} v_{u} \nonumber \\
&&-(A_{\kappa}\kappa)_{i}\text{Re}(e^{i\phi_{\upsilon_{\nu_{i}^c}}} e^{i\phi_{\upsilon_{\nu_{i}^c}}} e^{i\phi_{\upsilon_{\nu_{i}^c}}}) \upsilon_{\nu_{i}^c} +\Delta_{(2+i)(7+i)}, \\
M_{(\tilde{\nu}_{i}^{c})^{\text{Re}}(\tilde{\nu}_{j}^{c})^{\text{Im}}}^{2}\!\!\!\!\!&&=\Delta_{(2+i)(7+i)}.
\end{eqnarray}
\begin{eqnarray}
\Delta_{16}^{t}\!\!\!\!\!&&={{3G_{F}m_{t}^{4}}\over{2\sqrt{2}\pi^2\sin^2\beta}} {{|\mu|^2\cot\beta-\text{Re}(e^{i\phi_{\lambda}} e^{i\phi_{A_{t}}} e^{i\phi_{v_{u}}} e^{-i\phi_{v_d}} e^{i\phi_{\upsilon_{\nu_{i}^c}}}) A_{t}\mu   }\over{ m_{\tilde{t}_{1}}^2-m_{\tilde{t}_{2}}^2 }} \nonumber \\
&& \times {{-\text{Im}(e^{i\phi_{\lambda}} e^{i\phi_{A_{t}}} e^{i\phi_{v_{u}}} e^{-i\phi_{v_d}} e^{i\phi_{\upsilon_{\nu_{i}^c}}}) A_{t}\mu  }\over{ m_{\tilde{t}_{1}}^2-m_{\tilde{t}_{2}}^2 }} g(m_{\tilde{t}_{1}}^2,m_{\tilde{t}_{2}}^2), \\
\Delta_{17}^{t}\!\!\!\!\!&&={{3G_{F}m_{t}^{4}}\over{2\sqrt{2}\pi^2\sin^2\beta}} {{|\mu|^2\cot\beta-\text{Re}(e^{i\phi_{\lambda}} e^{i\phi_{A_{t}}} e^{i\phi_{v_{u}}} e^{-i\phi_{v_d}} e^{i\phi_{\upsilon_{\nu_{i}^c}}}) A_{t}\mu   }\over{ m_{\tilde{t}_{1}}^2-m_{\tilde{t}_{2}}^2 }} \nonumber \\
&& \times {{\text{Im}(e^{i\phi_{\lambda}} e^{i\phi_{A_{t}}} e^{i\phi_{v_{u}}} e^{-i\phi_{v_d}} e^{i\phi_{\upsilon_{\nu_{i}^c}}}) A_{t}\mu\tan\beta }\over{ m_{\tilde{t}_{1}}^2-m_{\tilde{t}_{2}}^2 }} g(m_{\tilde{t}_{1}}^2,m_{\tilde{t}_{2}}^2), \\
\Delta_{26}^{t}\!\!\!\!\!&&={{3G_{F}m_{t}^{4}}\over{2\sqrt{2}\pi^2\sin^2\beta}} {{-\text{Im}(e^{i\phi_{\lambda}} e^{i\phi_{A_{t}}} e^{i\phi_{v_{u}}} e^{-i\phi_{v_d}} e^{i\phi_{\upsilon_{\nu_{i}^c}}}) A_{t}\mu  }\over{ m_{\tilde{t}_{1}}^2-m_{\tilde{t}_{2}}^2 }} \nonumber \\
&&\times \Bigg\{\ln{m_{\tilde{t}_{1}}^2 \over m_{\tilde{t}_{2}}^2}+ {{|A_{t}|^2-\text{Re}(e^{i\phi_{\lambda}} e^{i\phi_{A_{t}}} e^{i\phi_{v_{u}}} e^{-i\phi_{v_d}} e^{i\phi_{\upsilon_{\nu_{i}^c}}}) A_{t}\mu\cot\beta }\over{ m_{\tilde{t}_{1}}^2-m_{\tilde{t}_{2}}^2 }} g(m_{\tilde{t}_{1}}^2,m_{\tilde{t}_{2}}^2)   \Bigg\}, \\
\Delta_{27}^{t}\!\!\!\!\!&&={{3G_{F}m_{t}^{4}}\over{2\sqrt{2}\pi^2\sin^2\beta}} {{\text{Im}(e^{i\phi_{\lambda}} e^{i\phi_{A_{t}}} e^{i\phi_{v_{u}}} e^{-i\phi_{v_d}} e^{i\phi_{\upsilon_{\nu_{i}^c}}}) A_{t}\mu\cot\beta  }\over{ m_{\tilde{t}_{1}}^2-m_{\tilde{t}_{2}}^2 }} \nonumber \\
&&\times \Bigg\{\ln{m_{\tilde{t}_{1}}^2 \over m_{\tilde{t}_{2}}^2}+ {{|A_{t}|^2-\text{Re}(e^{i\phi_{\lambda}} e^{i\phi_{A_{t}}} e^{i\phi_{v_{u}}} e^{-i\phi_{v_d}} e^{i\phi_{\upsilon_{\nu_{i}^c}}}) A_{t}\mu\cot\beta }\over{ m_{\tilde{t}_{1}}^2-m_{\tilde{t}_{2}}^2 }} g(m_{\tilde{t}_{1}}^2,m_{\tilde{t}_{2}}^2)   \Bigg\}, \\
\Delta_{1(7+i)}^{t}\!\!\!\!\!&&={{3G_{F}m_{t}^{4}}\over{2\sqrt{2}\pi^2\sin^2\beta}} {{|\mu|^2\cot\beta-\text{Re}(e^{i\phi_{\lambda}} e^{i\phi_{A_{t}}} e^{i\phi_{v_{u}}} e^{-i\phi_{v_d}} e^{i\phi_{\upsilon_{\nu_{i}^c}}})A_{t}\mu  }\over{ m_{\tilde{t}_{1}}^2-m_{\tilde{t}_{2}}^2 }}  \nonumber \\
&&\times {{-\text{Im}(e^{i\phi_{\lambda}} e^{i\phi_{A_{t}}} e^{i\phi_{v_{u}}} e^{-i\phi_{v_d}} e^{i\phi_{\upsilon_{\nu_{i}^c}}}) A_{t}\lambda_{i}v_{d}  }\over{m_{\tilde{t}_{1}}^2-m_{\tilde{t}_{2}}^2}}g(m_{\tilde{t}_{1}}^2,m_{\tilde{t}_{2}}^2), \\
\Delta_{2(7+i)}^{t}\!\!\!\!\!&&={{3G_{F}m_{t}^{4}}\over{2\sqrt{2}\pi^2\sin^2\beta}} {{-\text{Im}(e^{i\phi_{\lambda}} e^{i\phi_{A_{t}}} e^{i\phi_{v_{u}}} e^{-i\phi_{v_d}} e^{i\phi_{\upsilon_{\nu_{i}^c}}}) A_{t}\lambda_{i}v_{d}  }\over{m_{\tilde{t}_{1}}^2-m_{\tilde{t}_{2}}^2}} \nonumber \\
&&\times \Bigg\{\ln{m_{\tilde{t}_{1}}^2 \over m_{\tilde{t}_{2}}^2}+ {{|A_{t}|^2-\text{Re}(e^{i\phi_{\lambda}} e^{i\phi_{A_{t}}} e^{i\phi_{v_{u}}} e^{-i\phi_{v_d}} e^{i\phi_{\upsilon_{\nu_{i}^c}}}) A_{t}\mu\cot\beta }\over{ m_{\tilde{t}_{1}}^2-m_{\tilde{t}_{2}}^2 }} g(m_{\tilde{t}_{1}}^2,m_{\tilde{t}_{2}}^2)   \Bigg\}, \\
\Delta_{(2+i)6}^{t}\!\!\!\!\!&&={{3G_{F}m_{t}^{4}}\over{2\sqrt{2}\pi^2\sin^2\beta}} {{-\text{Im}(e^{i\phi_{\lambda}} e^{i\phi_{A_{t}}} e^{i\phi_{v_{u}}} e^{-i\phi_{v_d}} e^{i\phi_{\upsilon_{\nu_{i}^c}}}) A_{t}\mu  }\over{m_{\tilde{t}_{1}}^2-m_{\tilde{t}_{2}}^2}} \nonumber \\
&&\times {{{1\over2} \upsilon_{\nu_{i}^c} v_{d}(\lambda_{i}\lambda_{j}^{*}+\lambda_{i}^{*}\lambda_{j})-\text{Re}(e^{i\phi_{\lambda}} e^{i\phi_{A_{t}}} e^{i\phi_{v_{u}}} e^{-i\phi_{v_d}} e^{i\phi_{\upsilon_{\nu_{i}^c}}})A_{t}\lambda_{i}v_{d}  }\over{ m_{\tilde{t}_{1}}^2-m_{\tilde{t}_{2}}^2 }} g(m_{\tilde{t}_{1}}^2,m_{\tilde{t}_{2}}^2), \\
\Delta_{(2+i)7}^{t}\!\!\!\!\!&&={{3G_{F}m_{t}^{4}}\over{2\sqrt{2}\pi^2\sin^2\beta}} {{\text{Im}(e^{i\phi_{\lambda}} e^{i\phi_{A_{t}}} e^{i\phi_{v_{u}}} e^{-i\phi_{v_d}} e^{i\phi_{\upsilon_{\nu_{i}^c}}}) A_{t}\mu \cot\beta }\over{m_{\tilde{t}_{1}}^2-m_{\tilde{t}_{2}}^2}} g(m_{\tilde{t}_{1}}^2,m_{\tilde{t}_{2}}^2) \nonumber \\
&&\times {{{1\over2} \upsilon_{\nu_{i}^c} v_{d}(\lambda_{i}\lambda_{j}^{*}+\lambda_{i}^{*}\lambda_{j})-\text{Re}(e^{i\phi_{\lambda}} e^{i\phi_{A_{t}}} e^{i\phi_{v_{u}}} e^{-i\phi_{v_d}} e^{i\phi_{\upsilon_{\nu_{i}^c}}})A_{t}\lambda_{i}v_{d}  }\over{ m_{\tilde{t}_{1}}^2-m_{\tilde{t}_{2}}^2 }} , \\
\Delta_{(2+i)(7+i)}^{t}\!\!\!\!\!&&={{3G_{F}m_{t}^{4}}\over{2\sqrt{2}\pi^2\sin^2\beta}} {{-\text{Im}(e^{i\phi_{\lambda}} e^{i\phi_{A_{t}}} e^{i\phi_{v_{u}}} e^{-i\phi_{v_d}} e^{i\phi_{\upsilon_{\nu_{i}^c}}}) A_{t}\lambda_{i}v_{d} }\over{m_{\tilde{t}_{1}}^2-m_{\tilde{t}_{2}}^2}} g(m_{\tilde{t}_{1}}^2,m_{\tilde{t}_{2}}^2) \nonumber \\
&&\times {{{1\over2} \upsilon_{\nu_{i}^c} v_{d}(\lambda_{i}\lambda_{j}^{*}+\lambda_{i}^{*}\lambda_{j})-\text{Re}(e^{i\phi_{\lambda}} e^{i\phi_{A_{t}}} e^{i\phi_{v_{u}}} e^{-i\phi_{v_d}} e^{i\phi_{\upsilon_{\nu_{i}^c}}})A_{t}\lambda_{i}v_{d}  }\over{ m_{\tilde{t}_{1}}^2-m_{\tilde{t}_{2}}^2 }}, \\
\Delta_{16}^{b}\!\!\!\!\!&&={{3G_{F}m_{b}^{4}}\over{2\sqrt{2}\pi^2\cos^2\beta}} {{-\text{Im}(e^{i\phi_{\lambda}} e^{i\phi_{v_{u}}} e^{-i\phi_{v_d}} e^{i\phi_{\upsilon_{\nu_{i}^c}}})A_{b}\mu\tan\beta }\over{ m_{\tilde{b}_{1}}^2-m_{\tilde{b}_{2}}^2 }} \nonumber \\
&&\times \Bigg\{\ln{m_{\tilde{b}_{1}}^2\over m_{\tilde{b}_{2}}^2}+{{|A_{b}|^2-\text{Re}(e^{i\phi_{\lambda}} e^{i\phi_{v_{u}}} e^{-i\phi_{v_d}} e^{i\phi_{\upsilon_{\nu_{i}^c}}}) A_{b}\mu\tan\beta }\over{ m_{\tilde{b}_{1}}^2-m_{\tilde{b}_{2}}^2 }} g(m_{\tilde{b}_{1}}^2,m_{\tilde{b}_{2}}^2)    \Bigg\}, \\
\Delta_{17}^{b}\!\!\!\!\!&&={{3G_{F}m_{b}^{4}}\over{2\sqrt{2}\pi^2\cos^2\beta}} {{\text{Im}(e^{i\phi_{\lambda}} e^{i\phi_{v_{u}}} e^{-i\phi_{v_d}} e^{i\phi_{\upsilon_{\nu_{i}^c}}})A_{b}\mu }\over{ m_{\tilde{b}_{1}}^2-m_{\tilde{b}_{2}}^2 }} g(m_{\tilde{b}_{1}}^2,m_{\tilde{b}_{2}}^2) \nonumber \\
&&\times \Bigg\{\ln{m_{\tilde{b}_{1}}^2\over m_{\tilde{b}_{2}}^2}+{{|A_{b}|^2-\text{Re}(e^{i\phi_{\lambda}} e^{i\phi_{v_{u}}} e^{-i\phi_{v_d}} e^{i\phi_{\upsilon_{\nu_{i}^c}}}) A_{b}\mu\tan\beta }\over{ m_{\tilde{b}_{1}}^2-m_{\tilde{b}_{2}}^2 }}  \Bigg\}, \\
\Delta_{26}^{b}\!\!\!\!\!&&={{3G_{F}m_{b}^{4}}\over{2\sqrt{2}\pi^2\cos^2\beta}} {{-\text{Im}(e^{i\phi_{\lambda}} e^{i\phi_{v_{u}}} e^{-i\phi_{v_d}} e^{i\phi_{\upsilon_{\nu_{i}^c}}})A_{b}\mu\tan\beta }\over{ m_{\tilde{b}_{1}}^2-m_{\tilde{b}_{2}}^2 }}  \nonumber \\
&&\times {{|\mu|^2\tan\beta-\text{Re}(e^{i\phi_{\lambda}} e^{i\phi_{v_{u}}} e^{-i\phi_{v_d}} e^{i\phi_{\upsilon_{\nu_{i}^c}}})A_{b}\mu  }\over{ m_{\tilde{b}_{1}}^2-m_{\tilde{b}_{2}}^2 }}g(m_{\tilde{b}_{1}}^2,m_{\tilde{b}_{2}}^2),\\
\Delta_{27}^{b}\!\!\!\!\!&&={{3G_{F}m_{b}^{4}}\over{2\sqrt{2}\pi^2\cos^2\beta}} {{\text{Im}(e^{i\phi_{\lambda}} e^{i\phi_{v_{u}}} e^{-i\phi_{v_d}} e^{i\phi_{\upsilon_{\nu_{i}^c}}})A_{b}\mu }\over{ m_{\tilde{b}_{1}}^2-m_{\tilde{b}_{2}}^2 }} g(m_{\tilde{b}_{1}}^2,m_{\tilde{b}_{2}}^2) \nonumber \\
&&\times {{|\mu|^2\tan\beta-\text{Re}(e^{i\phi_{\lambda}} e^{i\phi_{v_{u}}} e^{-i\phi_{v_d}} e^{i\phi_{\upsilon_{\nu_{i}^c}}})A_{b}\mu  }\over{ m_{\tilde{b}_{1}}^2-m_{\tilde{b}_{2}}^2 }} ,\\
\Delta_{1(7+i)}^{b}\!\!\!\!\!&&={{3G_{F}m_{b}^{4}}\over{2\sqrt{2}\pi^2\cos^2\beta}} {{\text{Im}(e^{i\phi_{\lambda}} e^{i\phi_{v_{u}}} e^{-i\phi_{v_d}} e^{i\phi_{\upsilon_{\nu_{i}^c}}})A_{b}\lambda_{i}v_{u} }\over{ m_{\tilde{b}_{1}}^2-m_{\tilde{b}_{2}}^2 }} g(m_{\tilde{b}_{1}}^2,m_{\tilde{b}_{2}}^2) \nonumber \\
&&\times \Bigg\{\ln{m_{\tilde{b}_{1}}^2\over m_{\tilde{b}_{2}}^2}+{{|A_{b}|^2-\text{Re}(e^{i\phi_{\lambda}} e^{i\phi_{v_{u}}} e^{-i\phi_{v_d}} e^{i\phi_{\upsilon_{\nu_{i}^c}}}) A_{b}\mu\tan\beta }\over{ m_{\tilde{b}_{1}}^2-m_{\tilde{b}_{2}}^2 }}  \Bigg\} \\
\Delta_{2(7+i)}^{b}\!\!\!\!\!&&={{3G_{F}m_{b}^{4}}\over{2\sqrt{2}\pi^2\cos^2\beta}} {{\text{Im}(e^{i\phi_{\lambda}} e^{i\phi_{v_{u}}} e^{-i\phi_{v_d}} e^{i\phi_{\upsilon_{\nu_{i}^c}}})A_{b}\lambda_{i}v_{u} }\over{ m_{\tilde{b}_{1}}^2-m_{\tilde{b}_{2}}^2 }} g(m_{\tilde{b}_{1}}^2,m_{\tilde{b}_{2}}^2) \nonumber \\
&&\times {{|\mu|^2\tan\beta-\text{Re}(e^{i\phi_{\lambda}} e^{i\phi_{v_{u}}} e^{-i\phi_{v_d}} e^{i\phi_{\upsilon_{\nu_{i}^c}}})A_{b}\mu  }\over{ m_{\tilde{b}_{1}}^2-m_{\tilde{b}_{2}}^2 }},\\
\Delta_{(2+i)6}^{b}\!\!\!\!\!&&={{3G_{F}m_{b}^{4}}\over{2\sqrt{2}\pi^2\cos^2\beta}} {{-\text{Im}(e^{i\phi_{\lambda}} e^{i\phi_{v_{u}}} e^{-i\phi_{v_d}} e^{i\phi_{\upsilon_{\nu_{i}^c}}})A_{b}\mu\tan\beta }\over{ m_{\tilde{b}_{1}}^2-m_{\tilde{b}_{2}}^2 }} g(m_{\tilde{b}_{1}}^2,m_{\tilde{b}_{2}}^2)  \nonumber \\
&&\times{{{1\over2} \upsilon_{\nu_{i}^c} v_{u}(\lambda_{i}\lambda_{j}^{*}+\lambda_{i}^{*}\lambda_{j})\tan\beta-\text{Re}(e^{i\phi_{\lambda}}  e^{i\phi_{v_{u}}} e^{-i\phi_{v_d}} e^{i\phi_{\upsilon_{\nu_{i}^c}}})A_{b}\lambda_{i}v_{u}  }\over{ m_{\tilde{b}_{1}}^2-m_{\tilde{b}_{2}}^2 }},\\
\Delta_{(2+i)7}^{b}\!\!\!\!\!&&={{3G_{F}m_{b}^{4}}\over{2\sqrt{2}\pi^2\cos^2\beta}} {{\text{Im}(e^{i\phi_{\lambda}} e^{i\phi_{v_{u}}} e^{-i\phi_{v_d}} e^{i\phi_{\upsilon_{\nu_{i}^c}}})A_{b}\mu }\over{ m_{\tilde{b}_{1}}^2-m_{\tilde{b}_{2}}^2 }} g(m_{\tilde{b}_{1}}^2,m_{\tilde{b}_{2}}^2)  \nonumber \\
&&\times {{{1\over2} \upsilon_{\nu_{i}^c} v_{u}(\lambda_{i}\lambda_{j}^{*}+\lambda_{i}^{*}\lambda_{j})\tan\beta-\text{Re}(e^{i\phi_{\lambda}}  e^{i\phi_{v_{u}}} e^{-i\phi_{v_d}} e^{i\phi_{\upsilon_{\nu_{i}^c}}})A_{b}\lambda_{i}v_{u}  }\over{ m_{\tilde{b}_{1}}^2-m_{\tilde{b}_{2}}^2 }},\\
\Delta_{(2+i)(7+j)}^{b}\!\!\!\!\!&&={{3G_{F}m_{b}^{4}}\over{2\sqrt{2}\pi^2\cos^2\beta}} {{\text{Im}(e^{i\phi_{\lambda}} e^{i\phi_{v_{u}}} e^{-i\phi_{v_d}} e^{i\phi_{\upsilon_{\nu_{i}^c}}})A_{b}\lambda_{i}v_{u} }\over{ m_{\tilde{b}_{1}}^2-m_{\tilde{b}_{2}}^2 }} g(m_{\tilde{b}_{1}}^2,m_{\tilde{b}_{2}}^2) \nonumber \\
&&\times {{{1\over2} \upsilon_{\nu_{i}^c} v_{u}(\lambda_{i}\lambda_{j}^{*}+\lambda_{i}^{*}\lambda_{j})\tan\beta-\text{Re}(e^{i\phi_{\lambda}}  e^{i\phi_{v_{u}}} e^{-i\phi_{v_d}} e^{i\phi_{\upsilon_{\nu_{i}^c}}})A_{b}\lambda_{i}v_{u}  }\over{ m_{\tilde{b}_{1}}^2-m_{\tilde{b}_{2}}^2 }}.
\end{eqnarray}

\end{document}